\documentclass[11pt]{article}
\usepackage[preprint,nonatbib]{neurips_2026}
\usepackage{hyperref} 
\usepackage{authblk} 
\usepackage{booktabs}
\usepackage{graphicx}
\usepackage{float}
\usepackage[caption=false,font=footnotesize]{subfig}
\usepackage{amsfonts}
\usepackage{multicol}
\usepackage{tabularx}
\usepackage{enumitem,amsmath,bm,amssymb}
\usepackage{pifont,makecell}
\usepackage{fontawesome5}
\usepackage{tikz}
\usepackage{CJKutf8} 

\usepackage{multirow}
\usepackage{cite}
\usepackage{comment}
\usepackage{threeparttable}

\newcommand{\tablecaptionsep}{%
  \par\vspace{10pt}%
  \setlength{\abovecaptionskip}{0pt}%
}

\usepackage{xcolor}

\title{Qwen-Audio-3.0-TTS: Freely Controllable and Highly Robust Speech Synthesis with Multi-Stage Training Paradigm}

\author{%
  \parbox{0.66\textwidth}{%
    \centering
    \textit{(Equal contribution; alphabetical by given name.)}\\
    Bajian Xiang, Cheng Wen, Han Zhao, Hao Wang, Haoxu Wang,
    Jiawei Jin, Jiayan Cui, Jie Chen, Mengxi Nie, Tianyu Zhao,
    Weiqin Li, Xiang Lv, Xiangang Li, Yang Xiang, Yang Zhou\par
  }%
}
\affil{%
  \makebox[\textwidth][c]{%
    \parbox{0.72\textwidth}{%
      \centering\normalsize\bfseries
      \raisebox{-0.20em}{\includegraphics[height=1.05em]{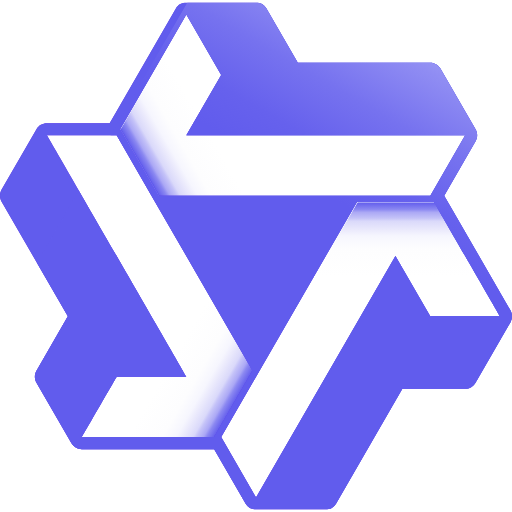}}%
      \enspace Alibaba Token Foundry\par
      \vspace{9pt}
      \normalfont\footnotesize\faVolumeUp\enspace
      \href{https://qwenaudio.github.io/FunAudioLLM.github.io/qwen-audio-3.0-tts/}{Demo Page}\par
    }%
  }%
}
\date{}
\begin{document}

\maketitle

\begin{abstract}

In this report, we present Qwen-Audio-3.0-TTS, a production-oriented speech synthesis system that jointly advances content consistency, speaker similarity, prosodic naturalness, audio quality, controllability, multilingual coverage, efficiency, and robustness. It combines a 12.5~Hz low-frame-rate speech tokenizer for reduced inference latency with a five-stage progressive training paradigm for coordinated language model (LM) and flow-matching model (FM) optimization. The model provides production-level control through free-style natural-language instructions and fine-grained inline tags, while supporting 16 languages, 20 Chinese dialect regions, one-pass long-form synthesis up to 3 minutes, and robust generation from noisy, reverberant, or unclear reference speech. Across SEED-TTS-Eval, CV3-Eval, instruction-following, long-form, and acoustic-robustness evaluations, Qwen-Audio-3.0-TTS achieves state-of-the-art performance on many reported dimensions or the strongest aggregate results. It also ranks first on the independent Artificial Analysis Text-to-Speech Leaderboard. These results establish Qwen-Audio-3.0-TTS as a strong foundation for production-level speech synthesis.

\end{abstract}

\section{Introduction}

Recent advances in large language models (LLMs), neural speech codecs, and diffusion/flow-based generative modeling have fundamentally reshaped text-to-speech (TTS) synthesis. Modern zero-shot systems learn from large-scale multi-speaker corpora and can reproduce the timbre and speaking characteristics of an unseen reference speaker without target-speaker fine-tuning \cite{DBLP:journals/corr/abs-2301-02111,DBLP:journals/corr/abs-2406-02430,DBLP:conf/nips/LeVSKSMWMAMH23,du2025cosyvoice3inthewildspeech}. The field has consequently moved beyond basic intelligibility toward robust multilingual and cross-lingual synthesis, fine-grained control of acoustic attributes, stable long-form generation, low-latency streaming, and resilience to adverse prompt conditions.

The modern in-context TTS landscape can be summarized by four overlapping paradigms. Autoregressive discrete-token systems, established by VALL-E and subsequently developed by Spark-TTS and Qwen3-TTS, integrate naturally with language modeling and enable low-latency causal generation, but quantization can discard fine acoustic information and decoding cost grows with token rate \cite{DBLP:journals/corr/abs-2301-02111,wang2025spark,hu2026qwen3ttstechnicalreport}. Non-autoregressive continuous systems such as Voicebox, E2~TTS, and F5-TTS instead synthesize continuous acoustic representations through parallel diffusion or flow matching, providing high fidelity at the cost of iterative utterance-level sampling that can complicate streaming \cite{DBLP:conf/nips/LeVSKSMWMAMH23,DBLP:journals/corr/abs-2406-18009,DBLP:journals/corr/abs-2410-06885}.

\begin{figure}[htbp]
\centering
\includegraphics[width=0.92\textwidth]{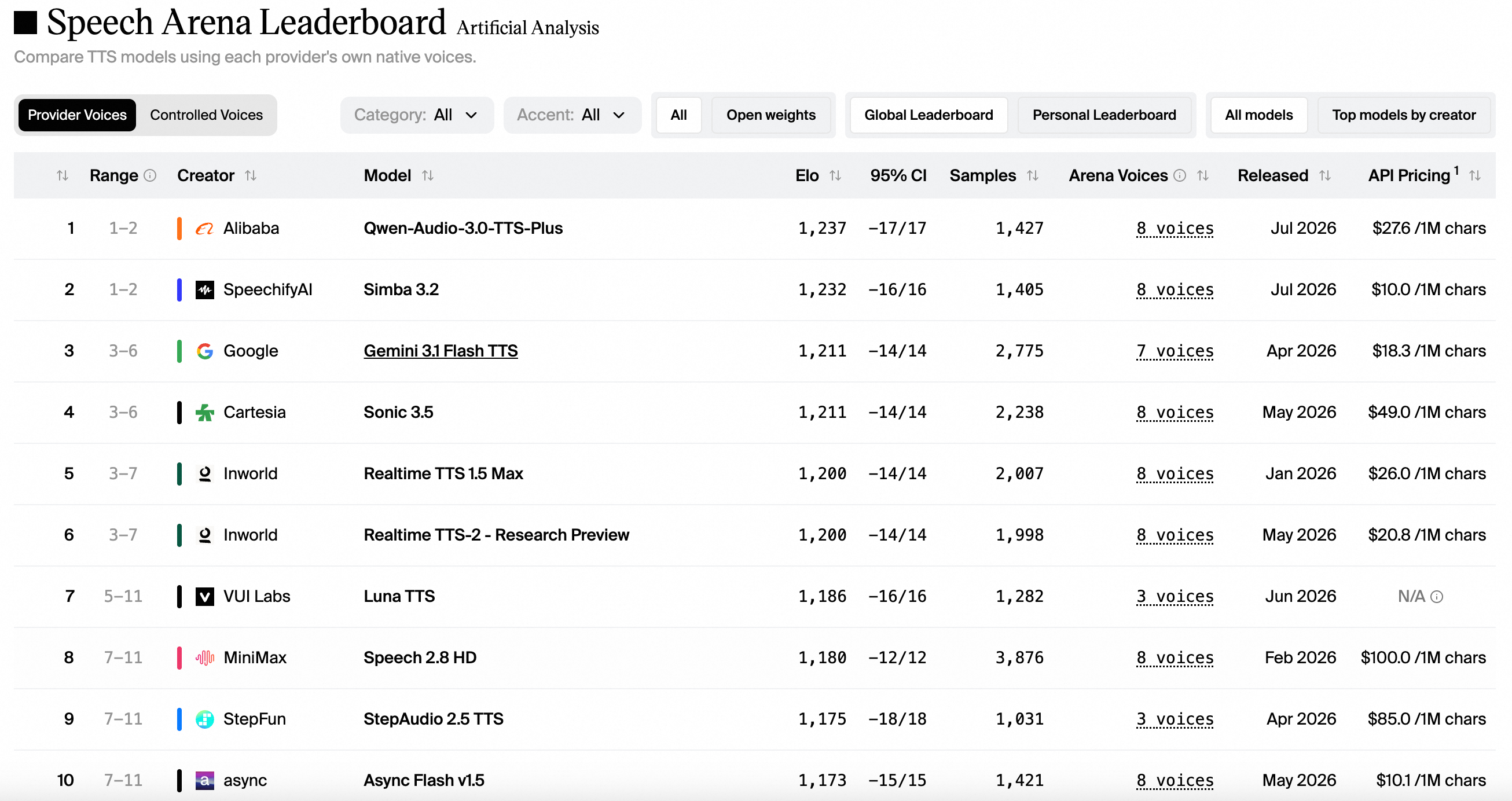}
\caption{Artificial Analysis Text-to-Speech Leaderboard snapshot on July 16, 2026.}
\label{fig:aa-arena}
\end{figure}

Hybrid systems then combined autoregressive semantic planning with continuous acoustic generation. Seed-TTS is an early representative, while CosyVoice introduced a supervised semantic speech tokenizer; CosyVoice2 improved codebook utilization and unified streaming and non-streaming synthesis, and CosyVoice3 strengthened in-the-wild generation through multi-task supervision, scaling, and post-training \cite{DBLP:journals/corr/abs-2406-02430,cosyvoice,du2024cosyvoice2,du2025cosyvoice3inthewildspeech}. This design separates linguistic planning from detailed rendering, but a discrete single-codebook interface remains an information and optimization bottleneck. More recently, continuous-autoregressive systems such as DiTAR, Dots.TTS, and VoxCPM2 have modeled continuous latents patch by patch without an external speech tokenizer \cite{jia2025ditar,lian2026dotsttstechnicalreport,zhou2026voxcpm2technicalreport}. They avoid quantization loss, but high-dimensional next-step generation, iterative local sampling, and error propagation can make synthesis stability and long-form consistency more sensitive to model and sampling design. These paradigms therefore offer complementary trade-offs rather than a single universally dominant solution.

The remaining challenge is to deliver these strengths simultaneously in a production-grade system. A practical TTS model must preserve content and speaker identity, generate clean, expressive, and natural audio, follow flexible control requests, cover diverse languages and dialects, stream with low latency, and remain stable with noisy, reverberant, or bandwidth-limited prompts. Standard short-form clean-speech benchmarks capture only part of these requirements and can obscure failures in multilingual, dialectal, long-form, and adverse acoustic conditions.

Qwen-Audio-3.0-TTS targets this complete quality--control--efficiency frontier in a single system. Building on CosyVoice2 and CosyVoice3, it retains efficient semantic planning while conditioning the flow-matching acoustic renderer on continuous LM hidden states and jointly optimizing the LM and FM, thereby alleviating the information bottleneck of a token-only interface. A 12.5~Hz tokenizer reduces autoregressive decoding cost, while high-quality data annealing, robustness training, and LM/FM reinforcement learning address content accuracy, prosodic naturalness, voice fidelity, perceptual quality, and adverse-prompt robustness. The resulting model combines production-level control and broad linguistic coverage with efficient, stable generation. Its key contributions are:

\begin{figure}[t]
\centering
\subfloat[Content Consistency.\label{fig:cv3-radar-content}]{%
  \includegraphics[width=0.485\textwidth]{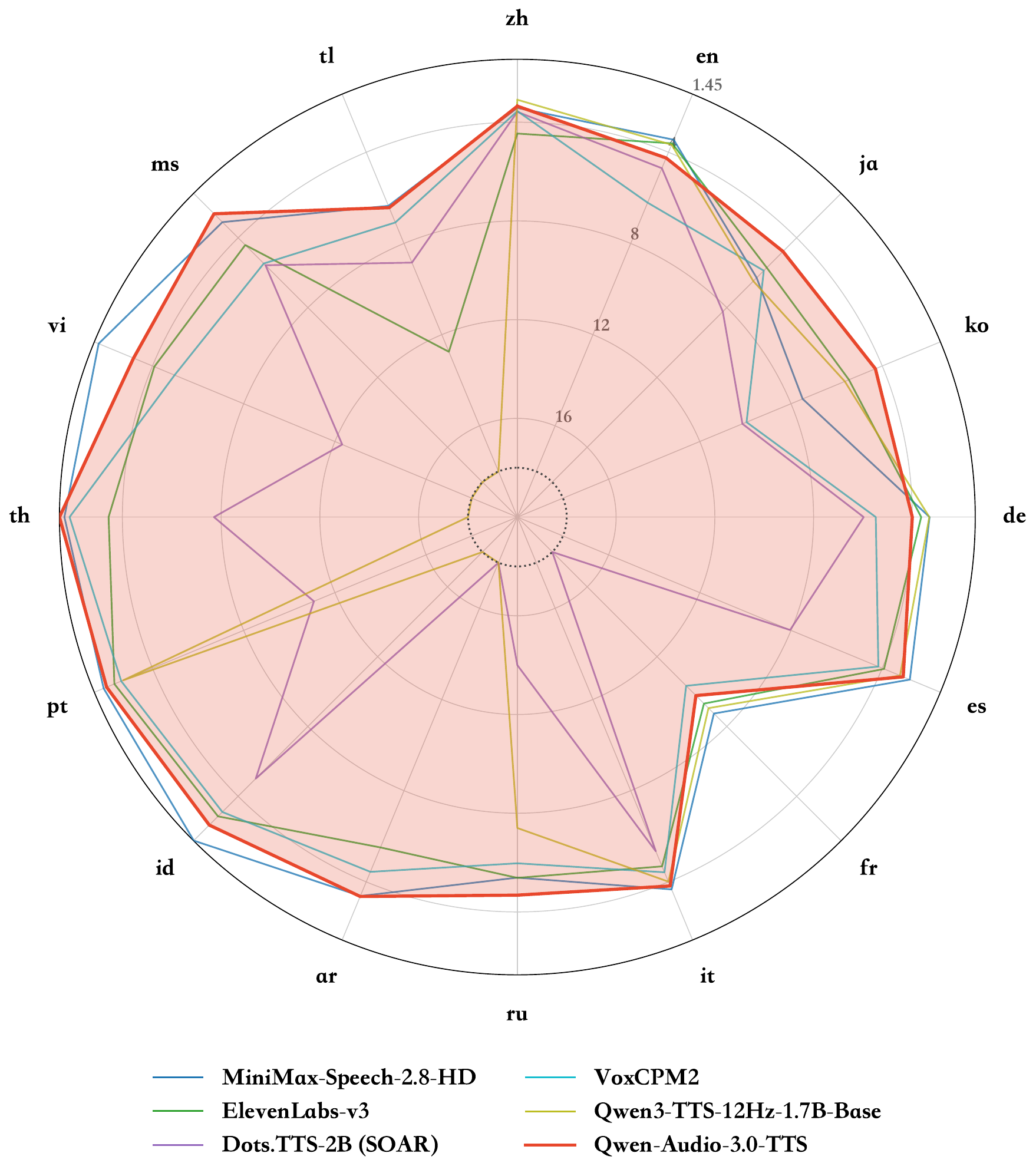}}
\hfill
\subfloat[Speaker Similarity.\label{fig:cv3-radar-sim}]{%
  \includegraphics[width=0.485\textwidth]{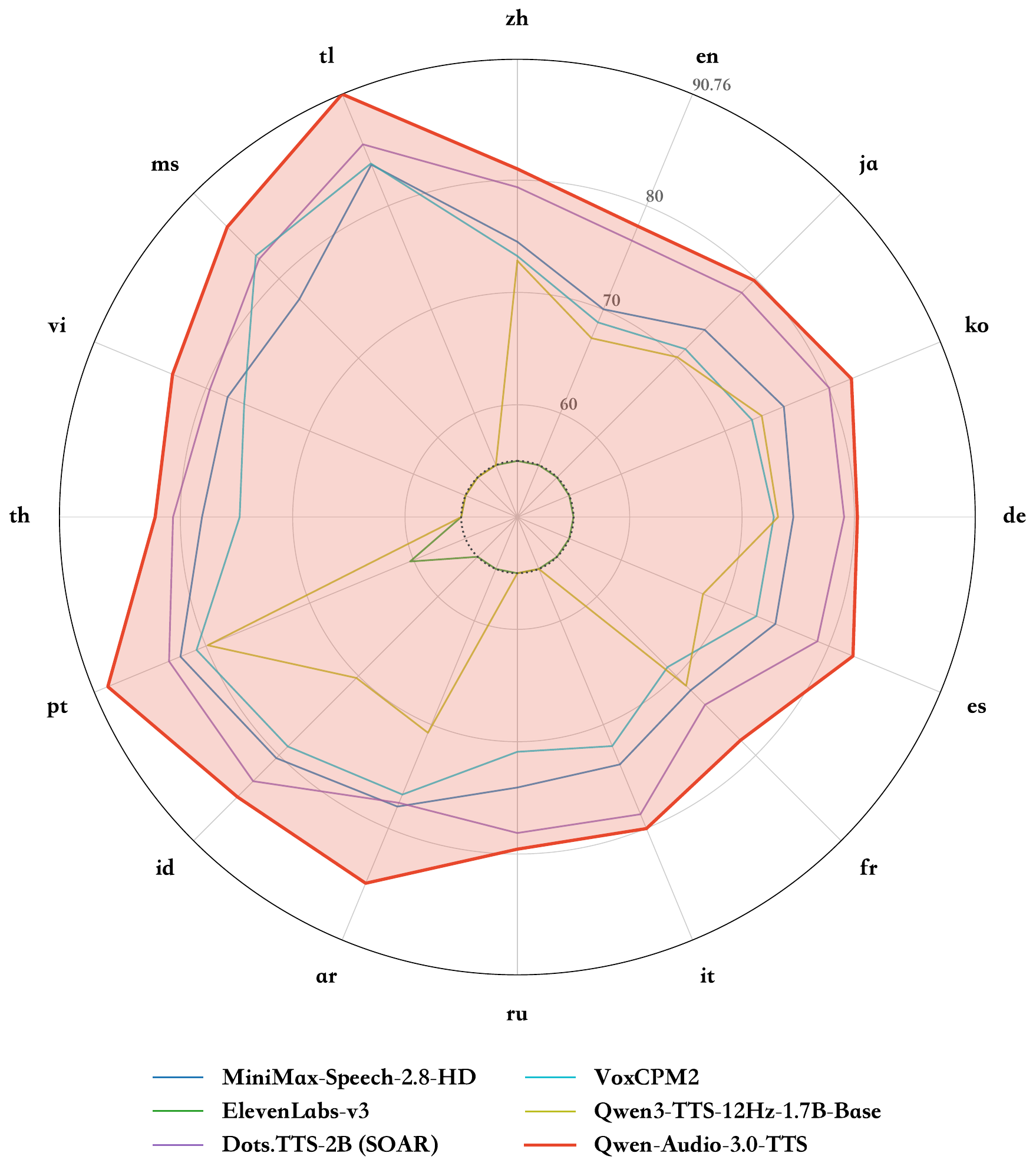}}
\caption{Multilingual comparison on the 16-language CV3-Eval benchmark. In both panels, farther from the center indicates better performance.}
\label{fig:cv3-radar}
\end{figure}

\begin{itemize}
\item \textbf{Low-frame-rate speech tokenizer:} A 12.5~Hz supervised speech tokenizer reduces autoregressive decoding cost while retaining content and speaker information.

\item \textbf{Progressive training paradigm:} The training pipeline combines independent LM and FM pretraining, joint training with high-quality data annealing, LM reinforcement learning, FM robustness training, and FM reinforcement learning to improve content consistency, prosodic naturalness, voice fidelity, perceptual quality, and robustness.

\item \textbf{Production-grade controllability:} The model interprets free-style natural-language instructions describing role, emotion, speaking style, rate, timbre, and accent. In parallel, 86 newly added fine-grained inline tags enable localized control at phrase and word level, including expressive transitions and non-verbal events such as laughter, breathing, coughing, and sighing.

\item \textbf{Broad and robust deployment coverage:} The model supports 16 languages, seven of them newly added, and 20 Chinese dialect regions; it handles hard text-normalization cases, one-pass synthesis up to 3 minutes, and degraded prompts without an explicit denoising mode. A two-stage speaker-adaptation protocol and vocoder super-resolution further support target-voice adaptation and 48~kHz output.

\item \textbf{Comprehensive evaluation:} We evaluate zero-shot voice cloning, multilingual and cross-lingual synthesis, free-style instruction following, fine-grained control, text normalization, long-form generation, adverse-prompt robustness, and 20-dialect synthesis through objective benchmarks and arena-style human evaluation.
\end{itemize}

Figure~\ref{fig:aa-arena} provides the snapshot of Artificial Analysis Text-to-Speech Arena leaderboard\footnote{\url{https://artificialanalysis.ai/text-to-speech/leaderboard/provider-voice?tab=leaderboard}} on July 16, 2026, which evaluates provider-native voices through blind pairwise preference tests with comparable gender and accent. The provider label \emph{Qwen-Audio-3.0-TTS-Plus} corresponds to the model reported as Qwen-Audio-3.0-TTS in this paper. Qwen-Audio-3.0-TTS-Plus ranks first with an Elo score of 1,237 from 1,427 samples. It has a displayed rank range of 1--2, and its 95\% confidence interval overlaps that of Simba~3.2; the leaderboard therefore places Qwen-Audio-3.0-TTS-Plus first by point estimate and within the statistically leading group.

In addition, extensive experiments demonstrate that Qwen-Audio-3.0-TTS has a favorable balance between content consistency, speaker similarity, prosodic naturalness, audio quality, and controllability. It achieves the best or highly competitive aggregate results on SEED-TTS-Eval, CV3-Eval, instruction-following, long-form synthesis, and adverse-prompt evaluation. It obtains the best aggregate free-style instruction-following scores in both Chinese and English, while standard-mode inference remains competitive with systems using explicit denoising. Figure~\ref{fig:cv3-radar} visualizes the per-language CV3-Eval comparison among several competitive providers, with content consistency in Figure~\ref{fig:cv3-radar-content} and speaker similarity in Figure~\ref{fig:cv3-radar-sim}.

\section{Qwen-Audio-3.0-TTS}

\begin{figure}[H]
    \centering
    \makebox[\textwidth][c]{%
        \includegraphics[
            width=1.0\textwidth,
            keepaspectratio
        ]{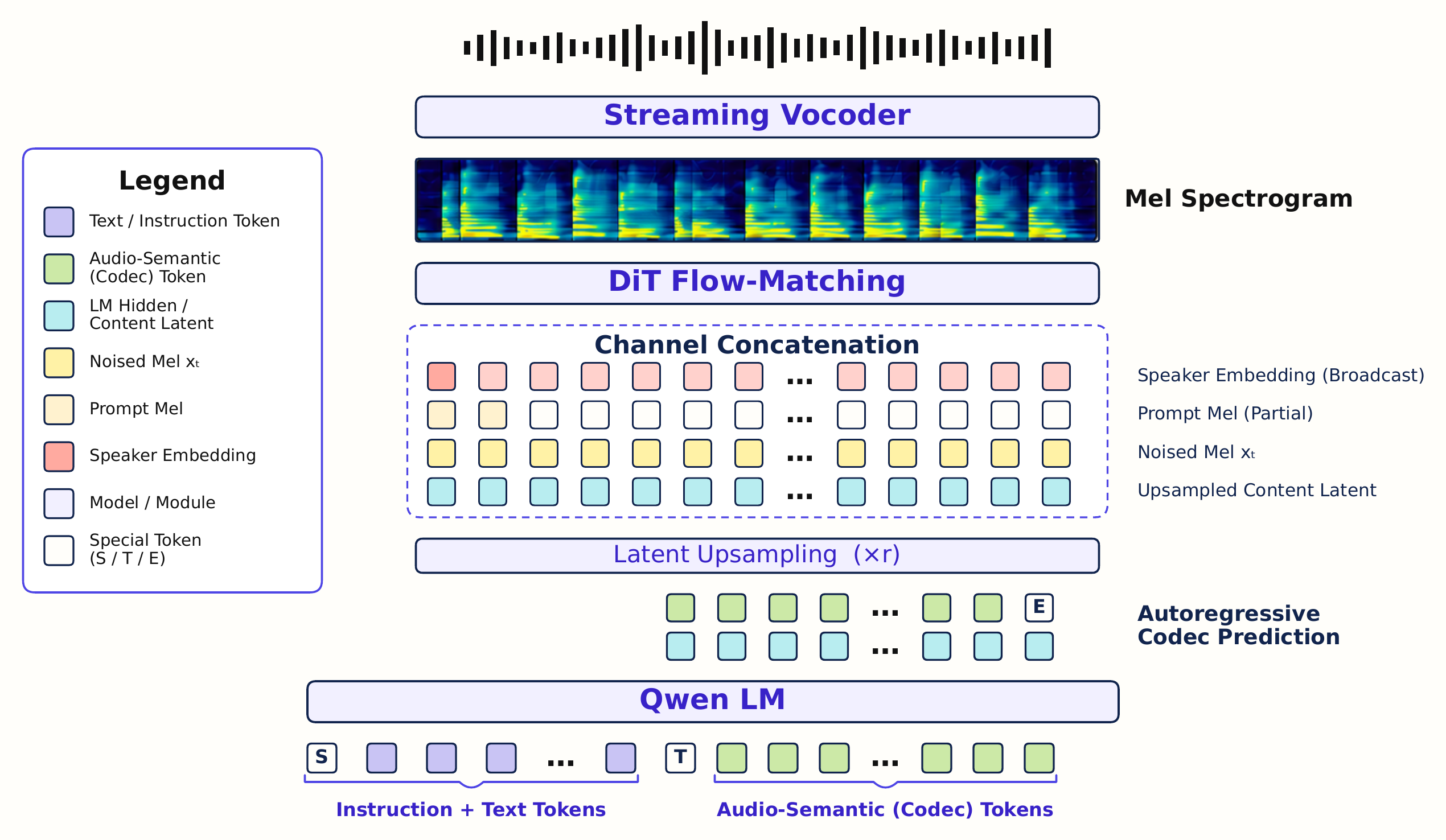}%
    }
    \caption{Overall architecture of Qwen-Audio-3.0-TTS, comprising the language model, flow-matching model, and vocoder.}
    \label{fig:architecture}
\end{figure}

As shown in Figure~\ref{fig:architecture}, Qwen-Audio-3.0-TTS is built on a three-component synthesis architecture comprising a language model (LM) for semantic token prediction, a flow-matching model (FM) for acoustic feature reconstruction, and a causal BigVGAN vocoder~\cite{lee2023bigvgan} for waveform synthesis. A 12.5~Hz low-frame-rate speech tokenizer reduces autoregressive decoding cost, while the progressive LM--FM training paradigm improves linguistic accuracy, acoustic fidelity, controllability, and robustness.

\subsection{Low-Frame-Rate Speech Tokenizer}

\begin{figure}[thb]
    \centering
    \includegraphics[width=0.70\textwidth]{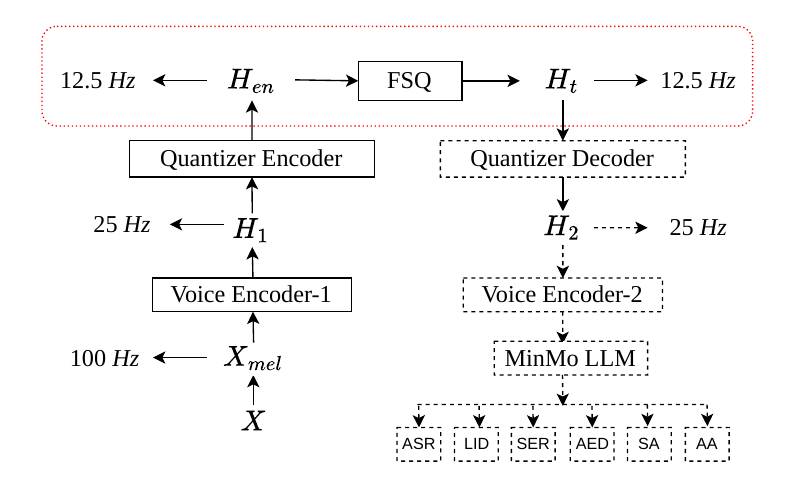}
    \caption{Architecture of the proposed supervised speech tokenizer. The tokenizer is optimized using a multi-task supervised learning objective encompassing automatic speech recognition (ASR), language identification (LID), speech emotion recognition (SER), audio event detection (AED), speaker analysis (SA), and broader audio analysis (AA) tasks.}
    \label{fig:tokenizer}
\end{figure}

As shown in Figure~\ref{fig:tokenizer}, the tokenizer follows the supervised design of CosyVoice3~\cite{du2025cosyvoice3inthewildspeech}: a causal SenseVoice encoder~\cite{funaduiollm} and Finite Scalar Quantization (FSQ)~\cite{DBLP:conf/iclr/MentzerMAT24} are integrated into a multi-task voice-encoder pipeline inspired by MinMo~\cite{chen2025minmo}. It maps Mel features to 12.5~Hz discrete tokens and learns the representation through supervised ASR, language, emotion, audio-event, speaker, and general audio-analysis tasks.

The encoder progressively downsamples the Mel sequence before quantization, and a corresponding decoder reconstructs an intermediate representation for multi-task supervision. This supervised bottleneck encourages the discrete tokens to retain linguistic content together with speaker, emotion, and acoustic-event information useful for speech generation. Training follows a continuous-to-quantized curriculum: the model first learns a stable continuous representation and subsequently activates FSQ to obtain discrete tokens.

Relative to CosyVoice3, we reduce the token rate from 25 to 12.5~Hz, substantially shortening the autoregressive sequence. A higher-capacity quantization space and broader audio-analysis supervision compensate for the stronger temporal compression, balancing generation efficiency with representation capacity.

\subsection{Multi-Stage Progressive Training Paradigm}

Qwen-Audio-3.0-TTS uses five progressive stages: independent LM and FM pretraining, joint LM--FM training with high-quality data annealing, LM reinforcement learning, FM robustness training, and FM reinforcement learning. Each stage starts from the preceding checkpoint and targets the capabilities most directly controlled by the corresponding module.

\subsubsection{Independent Pretraining of LM and FM}

The first stage follows the same decoupled LM--FM training recipe as our previous work, CosyVoice2~\cite{du2024cosyvoice2} and CosyVoice3~\cite{du2025cosyvoice3inthewildspeech}. The bi-streaming language model and the chunk-based flow-matching model are pretrained independently on large-scale, diverse speech data. The LM learns to predict the discrete semantic tokens produced by the speech tokenizer from the text and prompt context, thereby establishing robust content modeling and semantic planning capabilities. In parallel, the FM learns to reconstruct continuous acoustic features from tokenizer-derived discrete tokens, establishing a reliable mapping from quantized semantic representations to mel-spectrograms.

This independent pretraining provides a stable initialization for both components before they are coupled. It also preserves the modularity of the Cascade system: the LM can be scaled to improve linguistic and semantic modeling, while the FM can focus on acoustic fidelity and speaker reconstruction. The training data covers general speech, multilingual and dialect speech, and instruction-following data, providing broad coverage of languages, speakers, and speaking styles. The resulting LM and FM checkpoints together form the first-stage Cascade model and are used to initialize the joint-training stage described below.

\subsubsection{Joint LM-FM Training with High-Quality Data Annealing}

The second stage starts from the first-stage Cascade checkpoint and couples the pretrained LM and FM for end-to-end optimization. In alignment with the methodology of JoyVoice~\cite{yu2025joyvoice}, we condition the FM on continuous hidden states produced by the LM instead of discrete token embeddings. The semantic-token prediction path is retained, while the LM token-prediction objective and the FM flow-matching objective are optimized jointly. Consequently, the FM reconstruction loss can also shape the upstream LM representations through the shared hidden-state path.

This design reduces the information bottleneck introduced by discrete token quantization and mitigates the optimization mismatch between independently trained components. Compared with token IDs alone, the continuous LM hidden states preserve richer context that is useful for content realization, prosody, speaker characteristics, and instruction following. The FM can therefore exploit information that may not be fully represented by the discrete code sequence, while the token-prediction objective continues to provide a stable semantic learning signal. Instead of treating joint optimization as an isolated training setup in JoyVoice, our training schedule is progressive: it explicitly initializes joint training from the independently pretrained Cascade model.

Joint training first uses the broad-coverage data mixture to establish LM--FM alignment across languages, speakers, and styles. After this alignment has stabilized, training is annealed to a carefully curated high-quality subset containing cleaner and more expressive speech. Introducing this narrower distribution only in the later phase allows the model to retain the coverage learned from large-scale data while placing greater emphasis on acoustic fidelity, naturalness, expressiveness, and reliable instruction realization. Together, hidden-state conditioning and high-quality data annealing improve content consistency, prosodic detail, and end-to-end controllability.


\subsubsection{Language Model Reinforcement Learning}
\label{sec:lm_rl}

Starting from the jointly trained checkpoint, we optimize the autoregressive text-to-token LM while freezing the downstream FM and vocoder. Online Group Relative Policy Optimization (GRPO)~\cite{shao2024deepseekmath}, regularized by a KL penalty to a frozen
reference policy, compares groups of token rollouts under a composite reward that balances content consistency, duration robustness, generation diversity, and prosodic naturalness:
\begin{equation}
    R_{\mathrm{base},i}
    = \lambda_{\mathrm{content}}R_{\mathrm{content},i}
    + \lambda_{\mathrm{dur}}R_{\mathrm{dur},i}
    + \lambda_{\mathrm{div}}R_{\mathrm{div},i}
    + \lambda_{\mathrm{prosody}}R_{\mathrm{prosody},i}.
    \label{eq:base_reward}
\end{equation}
The content term is obtained from token-domain ASR; the duration term suppresses length outliers; the diversity term discourages mechanical collapse; and the prosody term rewards plausible alignment progression and pause timing. All rewards are computed before FM and vocoder inference,
enabling efficient token-only rollouts.

We additionally use a differentiable DiffRO branch~\cite{du2025cosyvoice3inthewildspeech} based on Gumbel--Softmax~\cite{jang2017categorical}.
To stabilize optimization, extreme anomalous rollouts, such as repetitions or missing stop tokens, are excluded from GRPO updates; DiffRO is further restricted to candidates with non-negative group-relative advantages:
\begin{equation}
    \mathcal{L}_{\mathrm{RL}}
    = \mathcal{L}_{\mathrm{GRPO}}
    + \lambda_{\mathrm{diff}}\mathcal{L}_{\mathrm{DiffRO}}^{+}.
    \label{eq:lm_rl_objective}
\end{equation}
GRPO supplies sequence-level relative preference, whereas DiffRO supplies selected token-level corrective gradients. LM reinforcement learning follows a two-phase curriculum: general generation optimization first excludes instruction-following, fine-grained-control, and dialect samples to avoid optimizing attributes not captured by the base reward; subsequent multi-task alignment adds dialect-classification correctness as an attribute reward,
improving dialect authenticity while preserving the general synthesis robustness acquired during the first phase.
With suitable attribute supervision, the same framework can be extended to instruction following and fine-grained control.
The resulting curriculum goes beyond WER-only optimization and balances
accuracy, naturalness, and controllability.

\subsubsection{Acoustic Robustness Training with Frozen LM}




Real-world prompts may be noisy, reverberant, bandwidth-limited, or recorded by low-quality devices. During the fourth stage, the LM is frozen and the FM is trained to recover clean, high-quality speech from degraded prompts while preserving timbre. The augmentation pool includes additive noise and reverberation; phone, Bluetooth, and laptop-microphone responses; far-field recording; physical blockage such as masks or hands over the microphone; codec, DAC, and amplifier artifacts; packet loss; strong echo; and compound settings such as noisy far-field meeting rooms or noise mixed with electronic distortion. Sampling these conditions during training integrates prompt enhancement into the cloning path rather than relying on a separate inference-time denoiser.

\subsubsection{Flow-Matching Reinforcement Learning}




The fifth stage applies FlowTTS-GRPO~\cite{flowtts_grpo, flowse_grpo, flowgrpo} to the FM, targeting speaker similarity and perceptual quality while the LM remains fixed. We convert deterministic ODE sampling $x_{t+\Delta t}=x_{t} + v_\theta(x_t,t)\Delta t$ into a marginal-preserving SDE sampler for on-policy exploration: 
\begin{equation}
x_{t+\Delta t} = x_{t,\mathrm{mean}} + \sigma_t\sqrt{\Delta t}\,\epsilon,
\qquad
\sigma_t = a\sqrt{\frac{1-t}{t}},
\quad \epsilon \sim \mathcal{N}(0,\mathbf{I}),
\label{eq:fm-rl-sde}
\end{equation}
\begin{equation}
x_{t,\mathrm{mean}} = x_t + \left[v_\theta(x_t,t)
+ \frac{\sigma_t^{2}}{2(1-t)}\left(-x_t+t\,v_\theta(x_t,t)\right)\right]\Delta t,
\label{eq:fm-rl-mean}
\end{equation}
 
where $v_\theta$ is the velocity field conditioned on LM hidden states, prompt mel features, and the speaker embedding, and $a$ controls exploration intensity. For each prompt, $G$ waveforms are sampled and the reward is normalized within the group: 
\begin{equation}
\hat{A}^{i} =
\frac{R(\hat{x}_1^{i},c)-\mathrm{mean}\{R(\hat{x}_1^{j},c)\}_{j=1}^{G}}
{\mathrm{std}\{R(\hat{x}_1^{j},c)\}_{j=1}^{G}}.
\label{eq:fm-rl-advantage}
\end{equation}

 where $\hat{x}_1^{i}$ is the $i$-th terminal waveform and $c$ contains its conditioning inputs. The reward combines speaker-verification similarity (SS), ASR intelligibility, and DNSMOS quality, each standardized by its per-batch standard deviation: 
\begin{equation}
R = \lambda_1 \frac{R_{\mathrm{SS}}}{\mathrm{std}(R_{\mathrm{SS}})} + \lambda_2 \frac{R_{\mathrm{ASR}}}{\mathrm{std}(R_{\mathrm{ASR}})} + \lambda_3 \frac{R_{\mathrm{MOS}}}{\mathrm{std}(R_{\mathrm{MOS}})},
\end{equation}

so that $\lambda_1$, $\lambda_2$, and $\lambda_3$ express the intended objective balance rather than raw reward variance. SDE exploration and policy optimization are restricted to an early-step window while later steps revert to the ODE, and classifier-free guidance~\cite{DBLP:journals/corr/abs-2207-12598} is omitted during training rollouts to widen exploration.

\subsection{Speaker Adaptation}

Speaker adaptation follows a two-stage supervised fine-tuning (SFT) procedure. Stage~1 jointly fine-tunes the LM and FM through chained adaptation rounds. In each round, the complete target-speaker set is paired with a refreshed replay subset matched by effective audio duration, maintaining broad linguistic and expressive coverage during adaptation. Stage~2 freezes the LM and refines the FM using target-speaker speech only, focusing the final update on speaker characteristics and local prosody.

A SFT-oriented super-resolution vocoder is trained to generate 48~kHz waveforms for richer harmonic detail and timbral expression. A multi-scale short-time Fourier transform discriminator supplies adversarial supervision at several time--frequency resolutions and reduces stripe-like high-frequency artifacts. Noise injected during training exposes the vocoder to imperfect upstream acoustic features and reduces the mismatch between ground-truth features used in training and predicted features encountered at inference.

\section{Experimental Settings}
\subsection{Speech Tokenizer}

The tokenizer follows the architecture in Section~2.1. It consumes 16~kHz audio through a Whisper-style frontend with 128 Mel-frequency bins, producing features at 100~Hz. Its causal SenseVoice encoder contains 32 Transformer layers with 1280 hidden dimensions and 20 attention heads. The initial 12-layer Voice Encoder-1 uses rotary positional embeddings (RoPE)~\cite{DBLP:journals/ijon/SuALPBL24} and downsamples the sequence to a 25~Hz representation $H_1$. A Quantizer Encoder then reduces both temporal and feature resolution to obtain $H_{en}$ at 12.5~Hz.

A 10-dimensional FSQ bottleneck, inserted after encoder layer 11, uses three levels per dimension and produces tokens $H_t$ from a codebook of $3^{10}=59{,}049$ entries. On the decoder side, a Quantizer Decoder upsamples the tokens to a 25~Hz representation $H_2$, which is processed by Voice Encoder-2 before entering the MinMo LLM. The language-model backbone used for supervised tokenizer training is initialized from \texttt{Qwen2.5-7B-Instruct}~\cite{qwen2.5}.

During continuous training, FSQ is bypassed; the tokenizer components are updated directly, while the language model is adapted with LoRA~\cite{hu2022lora}. During quantization training, FSQ is activated and the language-model weights are frozen. Both stages use cross-entropy objectives derived from the supervised tasks.

\subsection{Training Data of Qwen-Audio-3.0-TTS}

Qwen-Audio-3.0-TTS scales training data along five capability axes: multilingual and dialect coverage, free-style instruction following, fine-grained inline tags, long-form speech generation, and hard-case robustness. The model supports 16 languages, adding Malay, Tagalog, Arabic, Portuguese, Indonesian, Thai, and Vietnamese upon its predecessor CosyVoice3, and covers 20 Chinese dialect regions at finer geographic granularity.

Free-style instruction data covers speaker role, emotion, speaking style, rate, timbre, and accent. Fine-grained tags include localized controls for emotion, style, and speed as well as non-verbal events such as laughter, coughing, breathing, and sighing. Long-form speech is collected from narration-like settings and supports single-pass synthesis up to 3 minutes. Hard-case data covers polyphonic characters, rare and archaic characters, text-normalization numbers and symbols, and LaTeX mathematical expressions. During high-quality annealing, clean and expressive samples are emphasized.

\subsection{Evaluation Methods}

For evaluating Qwen-Audio-3.0-TTS's zero-shot speech generation capabilities, we focus on three key aspects: content consistency, speaker similarity, and audio quality. For content consistency, we measure the Character Error Rate (CER) or Word Error Rate (WER) of the ASR transcription against the given text, using Whisper-large V3 \cite{DBLP:conf/icml/RadfordKXBMS23} for English ASR and Paraformer \cite{DBLP:conf/interspeech/GaoZ0Y22,gao2023funasr} for Chinese ASR. To assess speaker similarity, we extract speaker embeddings from the generated speech using the ERes2Net speaker verification model~\cite{chen2023enhanced} and calculate the cosine similarity with the embedding of the reference speech. For audio quality, we score the generated speech using the DNSMOS network~\cite{DBLP:conf/icassp/ReddyGC22}, the scores of which show high correlations with human auditory perception.

Our core evaluations use \textbf{SEED-TTS-Eval}~\cite{DBLP:journals/corr/abs-2406-02430} and an extended \textbf{CV3-Eval}~\cite{du2025cosyvoice3inthewildspeech} covering seven additional languages. SEED-TTS-Eval reports both ERes2Net and WavLM similarity for comparison with prior work~\cite{chen2022large}.

We compare Qwen-Audio-3.0-TTS with widely used or competitive speech generation models. Non-autoregressive (NAR) baselines include F5-TTS \cite{DBLP:journals/corr/abs-2410-06885}, F5R-TTS \cite{sun2025f5r}, and LongCat-AudioDiT \cite{xin2026longcataudiodithighfidelitydiffusiontexttospeech}. Autoregressive (AR) baselines include Seed-TTS \cite{DBLP:journals/corr/abs-2406-02430}, FireRedTTS-2 \cite{xie2025fireredtts2}, IndexTTS2 \cite{zhou2025indextts2breakthroughemotionallyexpressive}, Qwen2.5-Omni \cite{xu2025qwen2}, Qwen3.5-Omni \cite{qwenteam2026qwen35omni}, Qwen3-TTS \cite{hu2026qwen3ttstechnicalreport}, Minimax-Speech \cite{minimax2025minimaxspeechintrinsiczeroshottexttospeech}, CosyVoice3 \cite{du2025cosyvoice3inthewildspeech}, Dots.TTS \cite{lian2026dotsttstechnicalreport}, and VoxCPM2 \cite{zhou2026voxcpm2technicalreport}.

Beyond these test sets, we introduce \textbf{Qwen-Audio-TTS-Eval}, a diagnostic benchmark for deployment-oriented speech generation, consisting of the following evaluation dimensions:

\begin{itemize}[leftmargin=*]

\item \textbf{Text Normalization:} 1,375 Chinese and English cases containing non-standard words, including numbers, dates, currencies, abbreviations, codes, formulas, and symbols, testing whether models can verbalize them correctly.

\item \textbf{Long-form Speech Generation:} 200 paragraph-level Chinese and English cases, typically producing utterances of 1.5--3 minutes, evaluating content consistency, speaker consistency, and prosodic stability in one-pass generation. English cases are adapted from~\cite{park2024librispeechlong} with same-speaker reference utterances, while Chinese cases are curated in-house.

\item \textbf{Acoustic Robustness:} 894 Chinese and English cases with noisy, reverberant, and unclear prompt speech, testing robustness to degraded enrollment audio.

\item \textbf{Instruction Following under Zero-shot Voice Cloning:} 440 cases covering single-attribute control of emotion, speech rate, and volume, as well as multi-attribute instructions expressed in natural language or structured key-value formats.

\end{itemize}

We also evaluate transfer from the pretrained model to speaker-fine-tuned models. Task-specific benchmark, calibration, and annotation protocols are reported alongside the corresponding results below.

\section{Experimental Results}
\subsection{Ablation of the Speech Tokenizer}

We investigate the impact of frame rate and codebook size on the tokenizer across automatic speech recognition (ASR) and downstream text-to-speech (TTS) tasks.

Intrinsic ASR results on Common Voice~\cite{commonvoice} and FLEURS~\cite{fleurs}, reported in Table~\ref{tab:tokenizer-asr}, show that increasing the codebook size recovers the performance loss caused by reducing the frame rate. Table~\ref{tab:tokenizer-tts} shows the same trend: reducing the frame rate from 25 to 12.5~Hz with the same 6,561-code vocabulary degrades content consistency and speaker similarity, whereas codebook scaling recovers the loss. Among the 12.5~Hz variants, the 59,049-code configuration achieves the best content consistency, while the 19,683-code configuration retains marginally higher speaker similarity, motivating the final accuracy--similarity--rate trade-off.

\begin{table}[htbp]
\caption{ASR error rates (\%) on Common Voice (CV) and FLEURS. CER is used for Chinese, Japanese, and Korean; WER is used for English. The best 12.5~Hz result is bold.}
\label{tab:tokenizer-asr}
\centering
\resizebox{\textwidth}{!}{%
\begin{tabular}{lcccccccc}
\toprule
\textbf{Tokenizer} & \textbf{Codebook} & \textbf{Rate} & \textbf{CV-zh} & \textbf{CV-en} & \textbf{CV-ja} & \textbf{CV-ko} & \textbf{FLEURS-zh} & \textbf{FLEURS-en} \\
\midrule
CosyVoice3 & 6,561 & 25 Hz & 10.63 & 13.07 & 15.61 & 11.35 & 3.77 & 5.43 \\
Qwen-Audio-3.0-TTS & 6,561 & 12.5 Hz & 11.23 & 15.40 & 18.68 & 13.22 & 4.18 & 5.33 \\
Qwen-Audio-3.0-TTS & 19,683 & 12.5 Hz & 10.79 & 13.39 & 16.63 & \textbf{11.45} & 4.00 & 4.91 \\
Qwen-Audio-3.0-TTS & 59,049 & 12.5 Hz & \textbf{10.24} & \textbf{12.52} & \textbf{15.21} & 11.70 & \textbf{3.85} & \textbf{4.69} \\
\bottomrule
\end{tabular}
}
\end{table}

\begin{table}[H]
\caption{Zero-shot TTS performance of different tokenizers on SEED-TTS-Eval. Content consistency is measured by CER/WER, and speaker similarity (SIM) is measured by ERes2Net and reported as a percentage. The best 12.5~Hz result in each column is shown in bold.}
\label{tab:tokenizer-tts}
\centering
\resizebox{\textwidth}{!}{%
\begin{tabular}{lcccccccc}
\toprule
\textbf{Tokenizer} & \textbf{Codebook} & \textbf{Frame Rate}
& \multicolumn{2}{c}{\textbf{test-zh}}
& \multicolumn{2}{c}{\textbf{test-en}}
& \multicolumn{2}{c}{\textbf{test-hard}} \\
\cmidrule(lr){4-5}\cmidrule(lr){6-7}\cmidrule(lr){8-9}
& \textbf{Size} & & \textbf{CER (\%)~$\downarrow$} & \textbf{SIM (\%)~$\uparrow$} & \textbf{WER (\%)~$\downarrow$} & \textbf{SIM (\%)~$\uparrow$} & \textbf{CER (\%)~$\downarrow$} & \textbf{SIM (\%)~$\uparrow$} \\
\midrule
CosyVoice3 & 6,561 & 25 Hz & 1.45 & 80.60 & 2.57 & 73.60 & 6.83 & 77.60 \\
Qwen-Audio-3.0-TTS & 6,561 & 12.5 Hz & 2.59 & 72.44 & 3.21 & 61.64 & 7.94 & 69.78 \\
Qwen-Audio-3.0-TTS & 19,683 & 12.5 Hz & 1.48 & \textbf{83.25} & 2.56 & \textbf{77.58} & 6.70 & \textbf{80.85} \\
Qwen-Audio-3.0-TTS & 59,049 & 12.5 Hz & \textbf{1.23} & 83.09 & \textbf{2.37} & 77.49 & \textbf{6.68} & 80.61 \\
\bottomrule
\end{tabular}
}
\end{table}

\subsection{Objective TTS Results on SEED-TTS-Eval}

\vspace{-6pt}

\begin{table*}[htb]
	\caption{
		Zero-shot TTS performance on SEED-TTS-Eval.
		Content consistency is measured by CER/WER, and speaker similarity
		(SIM) is reported as a cosine score.
		Values outside parentheses use WavLM and values inside parentheses
		use ERes2Net.
		\textbf{Bold} and \underline{underlined} values denote the best and second-best results
		in each column, respectively.
		$^\dagger$ ERes2Net SIM scores were computed by us using the publicly
		released models. Before reporting these scores, we verified that our
		reproduced CER/WER and WavLM SIM results closely matched those
		reported in the corresponding papers.
	}
    \vspace{3pt}
	\label{tab:tts-seed-test}
	\centering
	\setlength\tabcolsep{4.5pt}
	\resizebox{\textwidth}{!}{
		\begin{tabular}{lclclcl}
			\toprule
			\multirow{2}{*}{\textbf{Model}}
			& \multicolumn{2}{c}{\textbf{\emph{test-zh}}}
			& \multicolumn{2}{c}{\textbf{\emph{test-en}}}
			& \multicolumn{2}{c}{\textbf{\emph{test-hard}}} \\
			\cmidrule(r){2-3} \cmidrule(r){4-5} \cmidrule(r){6-7}
			& \textbf{CER (\%)~$\downarrow$}
			& \multicolumn{1}{c}{\textbf{SIM~$\uparrow$}}
			& \textbf{WER (\%)~$\downarrow$}
			& \multicolumn{1}{c}{\textbf{SIM~$\uparrow$}}
			& \textbf{CER (\%)~$\downarrow$}
			& \multicolumn{1}{c}{\textbf{SIM~$\uparrow$}} \\
			\midrule
			\textbf{Human}
			& 1.26 & 0.755~(0.775)
			& 2.14 & 0.734~(0.742)
			& - & \multicolumn{1}{c}{-} \\

			\textbf{Vocoder Resyn.}
			& 1.27 & 0.720
			& 2.17 & 0.700
			& - & \multicolumn{1}{c}{-} \\
			\midrule

			\multicolumn{7}{c}{\textbf{Non-autoregressive Models}} \\
			\midrule

			\textbf{F5-TTS (32 NFE)}~\cite{DBLP:journals/corr/abs-2410-06885}
			& 1.56 & 0.741~(0.794)
			& 1.83 & 0.647~(0.742)
			& 8.67 & 0.713~(0.762) \\

			\textbf{F5R-TTS}~\cite{sun2025f5r}
			& 1.37 & 0.754
			& - & \multicolumn{1}{c}{-}
			& 8.79 & 0.718 \\

			\textbf{LongCat-AudioDiT-3.5B}~\cite{xin2026longcataudiodithighfidelitydiffusiontexttospeech}
			& 1.09 & \textbf{0.818}~(0.806)$^\dagger$
			& 1.50 & \textbf{0.786}~(0.771)$^\dagger$
			& \underline{6.04} & \textbf{0.797}~(0.781)$^\dagger$ \\
			\midrule

			\multicolumn{7}{c}{\textbf{Autoregressive Models}} \\
			\midrule

			\textbf{Seed-TTS}~\cite{DBLP:journals/corr/abs-2406-02430}
			& 1.12 & 0.796
			& 2.25 & 0.762
			& 7.59 & 0.776 \\

			\textbf{FireRedTTS-2}~\cite{xie2025fireredtts2}
			& 1.14 & 0.736
			& 1.95 & 0.665
			& \multicolumn{1}{c}{-}
			& \multicolumn{1}{c}{-} \\

			\textbf{Qwen2.5-Omni-7B}~\cite{xu2025qwen2}
			& 1.70 & 0.752
			& 2.72 & 0.632
			& 7.97 & 0.747 \\


			\textbf{Qwen3.5-Omni-Plus}~\cite{qwenteam2026qwen35omni}
			& 0.99 & \multicolumn{1}{c}{-}
			& \underline{1.26} & \multicolumn{1}{c}{-}
			& \multicolumn{1}{c}{-}
			& \multicolumn{1}{c}{-} \\

			\textbf{Qwen3-TTS-12Hz-1.7B-Base}~\cite{hu2026qwen3ttstechnicalreport}
			& \textbf{0.77} & \multicolumn{1}{c}{-}
			& \textbf{1.24} & \multicolumn{1}{c}{-}
			& \multicolumn{1}{c}{-}
			& \multicolumn{1}{c}{-} \\

            \textbf{MiniMax-Speech}~\cite{minimax2025minimaxspeechintrinsiczeroshottexttospeech} & 0.99 & 0.799 & 1.90 & 0.738 
            & \multicolumn{1}{c}{-} & \multicolumn{1}{c}{-} \\

			\textbf{VoxCPM2}~\cite{zhou2026voxcpm2technicalreport}
			& 0.97 & 0.795~(0.756)$^\dagger$
			& 1.84 & 0.753~(0.725)$^\dagger$
			& 8.13 & 0.753~(0.704)$^\dagger$ \\

			\textbf{Dots.TTS-2B (SOAR)}~\cite{lian2026dotsttstechnicalreport}
			& 0.94 & \underline{0.810}~(0.818)$^\dagger$
			& 1.30 & \underline{0.771}~(\underline{0.792})$^\dagger$
			& 6.60 & \underline{0.795}~(0.800)$^\dagger$ \\

			\textbf{CosyVoice3-1.5B}~\cite{du2025cosyvoice3inthewildspeech}
			& 1.12 & 0.781~(\underline{0.837})
			& 2.21 & 0.720~(0.789)
			& \textbf{5.83} & 0.758~(\underline{0.816}) \\
			\midrule

			\textbf{Qwen-Audio-3.0-TTS}
			& \underline{0.84} & 0.792~(\textbf{0.847})
			& 1.54 & 0.762~(\textbf{0.815})
			& 7.00 & 0.768~(\textbf{0.824}) \\
			\bottomrule
		\end{tabular}
	}
\end{table*}

Table~\ref{tab:tts-seed-test} compares Qwen-Audio-3.0-TTS with recent state-of-the-art zero-shot TTS models. Content consistency is evaluated using CER/WER, and speaker similarity is measured by WavLM and ERes2Net. For methods marked with $^\dagger$, we compute the ERes2Net scores using their publicly released models after verifying that the reproduced CER/WER and WavLM results closely match those reported in the original papers.

Overall, Qwen-Audio-3.0-TTS achieves a strong balance between content accuracy and speaker similarity. It ranks second in CER on \emph{test-zh} while remaining competitive on \emph{test-en} and \emph{test-hard}. We find that pushing CER/WER lower through more aggressive optimization consistently comes at the expense of speech naturalness and expressiveness. Our model therefore targets a better overall trade-off instead of optimizing specifically for the lowest CER/WER. For speaker similarity, Qwen-Audio-3.0-TTS remains competitive under WavLM and achieves the highest ERes2Net scores across all three test sets. We also observe that WavLM and ERes2Net often produce different system rankings, suggesting that the two metrics capture complementary aspects of speaker similarity.

\subsection{Objective Evaluation on Multilingual Benchmark CV3-Eval}

\subsubsection{Results of Multilingual Voice Cloning}
\label{CV3-eval-multilingual}

We evaluate Qwen-Audio-3.0-TTS on the Multilingual Voice Cloning subset of CV3-Eval. Following the original CV3-Eval protocol, we extend the evaluation to several less commonly benchmarked languages, including Arabic (ar), Indonesian (id), Portuguese (pt), Thai (th), Vietnamese (vi), Malay (ms), and Tagalog (tl). We additionally compare against recent multilingual systems, including MiniMax-Speech-2.8-HD\footnote{\url{https://platform.minimax.io/docs/guides/models-intro}} and ElevenLabs-v3\footnote{\url{https://elevenlabs.io/docs/overview/models}} through their public APIs, as well as Dots.TTS-2B (SOAR), VoxCPM2, and Qwen3-TTS-12Hz-1.7B-Base using their publicly released open-source models. All systems are evaluated using the same CV3-Eval methodology. Table~\ref{tab:ml-clone} reports CER for Chinese, Japanese, and Korean, and WER for all other languages.

As shown in Table~\ref{tab:ml-clone}, Qwen-Audio-3.0-TTS achieves the best results in a broad range of languages, including Japanese, Korean, Russian, Arabic, Malay, and Thai, while remaining highly competitive on the others. Overall, the model demonstrates strong multilingual voice cloning performance across all 16 evaluated languages.

For the hard-zh and hard-en subsets in Table~\ref{tab:hard-ml-clone}, we additionally include LongCat-AudioDiT as a bilingual Chinese--English baseline. Under this challenging evaluation setting, Qwen-Audio-3.0-TTS achieves the best speaker similarity and DNSMOS on both subsets, while maintaining highly competitive WER performance. These results demonstrate its strong balance among intelligibility, speaker preservation, and perceptual quality.

\begin{table}[t]
    \caption{CER(\%) and WER(\%) on the CV3-Eval Multilingual Voice Cloning subset. MiniMax-Speech-2.8-HD and ElevenLabs-v3 are evaluated through their public APIs, while all other comparison systems use publicly released open-source models. Best scores are in \textbf{bold}, and second-best scores are \underline{underlined}. -- means the language is unsupported.}
    \label{tab:ml-clone}
    \centering{%
    \resizebox{\linewidth}{!}{%
    \setlength{\tabcolsep}{2.5pt}
    \begin{tabular}{lcccccccccccccccc}
    \toprule
    \textbf{Model} & \textbf{zh} & \textbf{en} & \textbf{ja} & \textbf{ko} & \textbf{de} & \textbf{es} & \textbf{fr} & \textbf{it} & \textbf{ru} & \textbf{ar} & \textbf{id} & \textbf{pt} & \textbf{th} & \textbf{vi} & \textbf{ms} & \textbf{tl} \\
    \midrule
    
    \multicolumn{17}{l}{\textbf{Commercial API Models}} \\
    
    MiniMax-Speech-2.8-HD
    & 3.42 & \textbf{3.45} & 6.29 & 7.49 & \textbf{3.30} & \textbf{2.79} & \textbf{8.74} & \textbf{3.67} & 5.39 & \underline{3.38} & \textbf{1.46} & \textbf{1.86} & \underline{1.65} & \textbf{1.64} & \underline{3.10} & \textbf{6.35} \\
    
    ElevenLabs-v3
    & 4.46 & \underline{3.61} & \underline{5.71} & \underline{5.46} & 3.64 & 3.92
    & 9.31 & 4.68 & \underline{5.38} & 5.51 & 2.85 & 2.33 & 3.44 & 4.08 & 4.41 & 12.8 \\
    
    \midrule
    \multicolumn{17}{l}{\textbf{Open-source Models}} \\
    
    Qwen3-TTS-12Hz-1.7B-Base
    & \textbf{3.09} & 3.67 & 6.48 & 5.64 & \underline{3.31} & 3.22
    & \underline{9.05} & 4.01 & 7.40 & -- & -- & 2.71 & -- & -- & -- & -- \\
    
    Dots.TTS-2B (SOAR)
    & 3.58 & 4.70 & 8.23 & 10.1 & 5.97 & 8.04 & 35.7 & 5.34 & 14.0 & -- & 5.01 & 11.1 & 7.71 & 12.3 & 5.57 & 8.84 \\
    
    VoxCPM2
    & 3.55 & 6.21 & 5.88 & 9.95 & 5.48 & 4.17 & 10.3 & 4.42 & 5.97 & 4.44 & 3.10 & 2.63 & 1.86 & 4.97 & 5.47 & 7.08 \\
    
    CosyVoice3-0.5B
    & 3.89 & 5.24 & 10.4 & 12.8 & 7.41 & 4.25 & 12.9 & 6.68 & 6.77 & -- & -- & -- & -- & -- & -- & -- \\
    
    CosyVoice3-1.5B
    & 3.91 & 4.99 & 7.57 & 5.69 & 6.43 & 4.47 & 11.8 & 10.5 & 6.64 & -- & -- & -- & -- & -- & -- & -- \\
    
    \midrule
    
    \textbf{Qwen-Audio-3.0-TTS}
    & \underline{3.35} & 4.25 & \textbf{4.78} & \textbf{4.30} & 4.00 & \underline{3.08}
    & 9.77 & \underline{3.82} & \textbf{4.68} & \textbf{3.36}
    & \underline{2.35} & \underline{1.99} & \textbf{1.45} & \underline{3.17}
    & \textbf{2.62} & \underline{6.43} \\
    
    \bottomrule
    \end{tabular}%
    }
    }
\end{table}

\begin{table}[t]
\caption{WER/CER, ERes2Net speaker similarity (SIM), and DNSMOS on the hard-zh and hard-en subsets of CV3-Eval. The best result in each column is shown in \textbf{bold}, and the second-best result is \underline{underlined}.}
\label{tab:hard-ml-clone}
\centering
\fontsize{9pt}{9pt}\selectfont
\resizebox{0.98\textwidth}{!}{%
\setlength{\tabcolsep}{5pt}
\begin{tabular}{lcccccc}
\toprule
\multirow{2}{*}{\textbf{Model}} & \multicolumn{3}{c}{\textbf{hard-zh}} & \multicolumn{3}{c}{\textbf{hard-en}} \\
\cmidrule(lr){2-4} \cmidrule(lr){5-7}
 & \textbf{CER (\%)~$\downarrow$} & \textbf{SIM (\%)~$\uparrow$} & \textbf{DNSMOS~$\uparrow$} & \textbf{WER (\%)~$\downarrow$} & \textbf{SIM (\%)~$\uparrow$} & \textbf{DNSMOS~$\uparrow$} \\
\midrule

\multicolumn{7}{l}{\textbf{Commercial API Models}}\\

MiniMax-Speech-2.8-HD & \textbf{7.42} & 74.6 & 3.74 & 7.37 & 72.1 & 3.80 \\
ElevenLabs-v3 & 10.66 & 50.3 & \underline{3.81} & \textbf{5.84} & 48.9 & 3.92 \\

\midrule

\multicolumn{7}{l}{\textbf{Open-source Models}}\\

Qwen3-TTS-12Hz-1.7B-Base & 11.24 & 69.1 & 3.79 & \underline{6.53} & 66.1 & 3.88 \\
LongCat-AudioDiT-3.5B & 9.24 & 72.8 & 3.75 & 8.50 & 73.9 & 3.84 \\
Dots.TTS-2B (SOAR) & 11.75 & 77.5 & 3.65 & 11.69 & 76.0 & 3.72 \\
VoxCPM2 & 8.10 & 69.9 & 3.63 & 7.48 & 67.0 & 3.73 \\
CosyVoice3-1.5B & 9.77 & \underline{78.5} & 3.79 & 10.55 & \underline{76.1} & \underline{3.95} \\

\midrule
\textbf{Qwen-Audio-3.0-TTS} & \underline{7.44} & \textbf{78.7} & \textbf{3.93} & 6.71 & \textbf{76.6} & \textbf{4.04} \\

\bottomrule
\end{tabular}%
}
\end{table}

\subsubsection{Results of Cross-lingual Voice Cloning}

Table~\ref{tab:cl-clone} reports the WER/CER results on the CV3-Eval Cross-lingual Voice Cloning subset, comparing recent commercial API systems and open-source models. For readability, a few substantially higher results are omitted and denoted by ``-'' in the table. Across all 12 transfer directions, Qwen-Audio-3.0-TTS achieves the best result in eight and the second-best result in the remaining four, consistently ranking among the strongest systems across all evaluated language pairs. It also outperforms CosyVoice3-1.5B in every direction and reduces the average error from 10.09\% to 4.05\%, a relative reduction of approximately 60\%, demonstrating strong cross-lingual stability across diverse source and target languages.

\begin{table}[H]
    \caption{WER/CER (\%, $\downarrow$) on the CV3-Eval Cross-lingual Voice Cloning subset. Top-level headers denote target languages and second-level headers denote source languages. Bold and underlined values indicate the best and second-best results in each transfer direction. For readability, entries with substantially higher WER/CER are omitted and denoted by ``--''.}
    \label{tab:cl-clone}
    \centering{%
    \resizebox{\textwidth}{!}{%
    \begin{tabular}{lcccccccccccc}
    \toprule
    \multirow{2}{*}{\textbf{Model}}  & \multicolumn{3}{c}{\textbf{to-zh}} & \multicolumn{3}{c}{\textbf{to-en}} & \multicolumn{3}{c}{\textbf{to-ja}} & \multicolumn{3}{c}{\textbf{to-ko}} \\ 
    \cmidrule(lr){2-4} \cmidrule(lr){5-7} \cmidrule(lr){8-10} \cmidrule(lr){11-13}
      & \textbf{en} & \textbf{ja} & \textbf{ko} & \textbf{zh} & \textbf{ja} & \textbf{ko} & \textbf{zh} & \textbf{en} & \textbf{ko} & \textbf{zh} & \textbf{en} & \textbf{ja} \\
    \midrule
    
    \multicolumn{7}{l}{\textbf{Commercial API Models}} \\
    MiniMax-Speech-2.8-HD & 9.96 & 6.07 & 3.63 & 3.59 & 5.79 & 4.28 & 30.7 & 13.7 & 6.39 & 6.45 & 6.78 & 11.3 \\
    ElevenLabs-v3 & 7.15 & 6.17 & 2.82 & 4.86 & 5.30 & 4.88 & 10.7 & 12.4 & 6.00 & 6.26 & 5.48 & 8.01 \\

    \midrule
    \multicolumn{13}{l}{\textbf{Open-source Models}} \\
    
    Qwen3-TTS-12Hz-1.7B-Base & \textbf{4.77} & \underline{3.43} & \textbf{1.08} & \underline{2.77} & \textbf{3.04} & \textbf{3.09} & \underline{8.40} & \underline{7.21} & \underline{3.67} & \underline{4.82} & \underline{5.14} & \underline{5.59} \\
    
    Dots.TTS-2B (SOAR) & 8.07 & -- & 2.61 & 4.31 & 8.02 & 4.74 & 16.1 & -- & -- & 12.4 & 18.6 & 13.5 \\
    VoxCPM2 & 7.76 & -- & 3.42 & 5.26 & 6.22 & 7.15 & -- & -- & -- & 5.73 & 10.4 & 10.6 \\
    CosyVoice3-1.5B & 8.01 & 6.78 & 3.30 & 4.32 & 5.39 & 5.94 & 13.7 & 13.4 & 4.19 & 31.6 & 14.0 & 10.5 \\
    \midrule
    \textbf{Qwen-Audio-3.0-TTS} & \underline{5.23} & \textbf{3.29} & \underline{1.09} & \textbf{2.40} & \underline{3.15} & \underline{3.54} & \textbf{6.53} & \textbf{6.66} & \textbf{2.98} & \textbf{4.27} & \textbf{4.34} & \textbf{5.15} \\
    
    \bottomrule
    \end{tabular}}
    }
\end{table}

\subsection{Objective TTS Results on Qwen-Audio-TTS-Eval}

\subsubsection{Results of Text Normalization Ability}

The benchmark contains five categories. \textbf{Num.} covers numbers, dates, and times (for example, ``2023-12-01'' and ``VII''); \textbf{Fin.} covers monetary and financial expressions (``€99.99'' and ``\$4.8M''); \textbf{Acr.} covers abbreviations, acronyms, and mixed readings (``U-lock'' and ``CRISPR''); \textbf{Code} covers serial numbers, codes, and addresses (``V3.2.1'' and ``sales-2024@ali.com''); and \textbf{Expr.} covers formulas, units, and symbols, including $P_t/P_r$ and $\Delta G=\Delta H-T\Delta S$.

Gemini-2.5-Pro~\cite{comanici2025gemini25} receives the original text, the synthesized-audio ASR transcript, a human-authored list of acceptable verbalizations, and the category-specific evaluation focus. It assigns a binary score to the target expression while disregarding ASR errors and discrepancies outside that focus. Scores are averaged within categories and over the complete benchmark.

Table~\ref{tab:tn-results-full} summarizes the overall and category-level results. Qwen-Audio-3.0-TTS achieves the best overall accuracy on both Chinese (68.7\%) and English (65.7\%) evaluation sets. It also shows competitive performance across different categories, demonstrating its ability to handle diverse text normalization scenarios.

\begin{table}[htbp]
    \caption{Complete category-level text-normalization accuracy. Higher is better; bold and underline denote the best and second-best results.}
    \label{tab:tn-results-full}
    \centering
    \resizebox{\textwidth}{!}{%
    \small
    \setlength{\tabcolsep}{5pt}
    \begin{tabular}{lcccccccccccc}
        \toprule
        \multirow{2}{*}{\textbf{Model}}
        & \multicolumn{2}{c}{\textbf{Overall}} & \multicolumn{2}{c}{\textbf{Num.}}
        & \multicolumn{2}{c}{\textbf{Fin.}} & \multicolumn{2}{c}{\textbf{Acr.}}
        & \multicolumn{2}{c}{\textbf{Code}} & \multicolumn{2}{c}{\textbf{Expr.}} \\
        \cmidrule(lr){2-3}\cmidrule(lr){4-5}\cmidrule(lr){6-7}
        \cmidrule(lr){8-9}\cmidrule(lr){10-11}\cmidrule(lr){12-13}
        & zh & en & zh & en & zh & en & zh & en & zh & en & zh & en \\
        \midrule
        Qwen3-TTS-12Hz-1.7B-Base & 57.0 & \underline{60.5} & 74.2 & \textbf{78.7} & \underline{40.5} & 63.0 & 41.5 & \textbf{63.3} & 73.9 & \underline{47.1} & \underline{24.8} & \underline{44.1} \\
        LongCat-AudioDiT-3.5B & 2.0 & 7.4 & 2.7 & 3.6 & 0.0 & 10.0 & 4.9 & 27.3 & 0.8 & 0.0 & 1.0 & 2.8 \\
        Dots.TTS-2B (SOAR) & 38.6 & 46.9 & 40.6 & 67.7 & 28.4 & 49.5 & 38.3 & 52.0 & 72.3 & 31.0 & 3.8 & 27.5 \\
        VoxCPM2 & 55.4 & 48.6 & 72.9 & 61.8 & 33.8 & 51.0 & 39.0 & 58.6 & \underline{82.4} & 42.0 & 15.2 & 25.2 \\
        CosyVoice3-1.5B & \underline{59.3} & 54.2 & \underline{81.3} & 70.2 & \underline{40.5} & \underline{77.0} & \underline{45.1} & 55.5 & 80.7 & 37.4 & 12.4 & 32.2 \\
        \midrule
        Qwen-Audio-3.0-TTS & \textbf{68.7} & \textbf{65.7} & \textbf{84.2} & \underline{78.2} & \textbf{43.7} & \textbf{82.8} & \textbf{50.6} & \underline{59.4} & \textbf{89.7} & \textbf{61.5} & \textbf{43.8} & \textbf{45.1} \\
        \bottomrule
    \end{tabular}
    }
\end{table}

\subsubsection{Results of Long-form Speech Generation}

\begin{table}[htbp]
    \caption{Complete one-pass long-form comparison by input-length bucket. Length is measured in characters for Chinese and words for English; duration is estimated from the Qwen-Audio-3.0-TTS outputs. ``--'' indicates a metric that is not applicable, while ``--$^\dagger$'' denotes omitted CER/WER values that fall substantially outside the effective comparison range under our evaluation setting.}
    \label{tab:long-form-tts-full}
    \centering
    \scriptsize
    \setlength{\tabcolsep}{1.2pt}
    \renewcommand{\arraystretch}{1.06}
    \resizebox{\textwidth}{!}{%
    \begin{tabular}{lcccccc cccccc}
        \toprule
        \multirow{2}{*}{\textbf{Model}}
        & \multicolumn{4}{c}{\textbf{zh CER (\%)~$\downarrow$}}
        & \multicolumn{2}{c}{\textbf{zh SIM~$\uparrow$}}
        & \multicolumn{4}{c}{\textbf{en WER (\%)~$\downarrow$}}
        & \multicolumn{2}{c}{\textbf{en SIM~$\uparrow$}} \\
        \cmidrule(lr){2-5}\cmidrule(lr){6-7}\cmidrule(lr){8-11}\cmidrule(lr){12-13}
        & \textbf{short} & \textbf{mid} & \textbf{long} & \textbf{all}
        & \textbf{P-SIM} & \textbf{S-SIM}
        & \textbf{short} & \textbf{mid} & \textbf{long} & \textbf{all}
        & \textbf{P-SIM} & \textbf{S-SIM} \\
        \midrule
        \textit{Text length ($\mu\pm\sigma$)}
        & 496$\pm$19 & 634$\pm$26 & 735$\pm$24 & 630$\pm$99 & -- & --
        & 248$\pm$26 & 345$\pm$31 & 451$\pm$29 & 358$\pm$88 & -- & -- \\
        \textit{Audio duration ($\mu\pm\sigma$, s)}
        & 116$\pm$9 & 146$\pm$11 & 161$\pm$23 & 142$\pm$24 & -- & --
        & 86$\pm$14 & 115$\pm$18 & 163$\pm$27 & 125$\pm$38 & -- & -- \\
        \textit{$N$ samples} & 29 & 35 & 36 & 100 & 100 & 100 & 29 & 32 & 39 & 100 & 100 & 100 \\
        \midrule
        Qwen3-TTS-12Hz-1.7B-Base & \underline{0.34} & 1.89 & 5.79 & 2.84 & 63.11 & 88.98 & \underline{3.04} & \textbf{4.25} & 6.57 & \underline{4.81} & 68.56 & 90.49 \\
        LongCat-AudioDiT-3.5B & --$^\dagger$ & --$^\dagger$ & --$^\dagger$ & --$^\dagger$ & 70.15 & 87.10 & --$^\dagger$ & --$^\dagger$ & --$^\dagger$ & --$^\dagger$ & 71.07 & 88.24 \\
        Dots.TTS-2B (SOAR) & 16.66 & 36.03 & 47.31 & 34.47 & 78.47 & 89.74 & 13.42 & 20.67 & 47.85 & 29.17 & 81.80 & 91.48 \\
        VoxCPM2 & 0.54 & \underline{0.58} & \textbf{0.49} & \textbf{0.54} & 61.73 & 86.80 & \textbf{2.33} & \textbf{4.25} & \textbf{2.98} & \textbf{3.20} & 68.95 & 90.39 \\
        CosyVoice3-1.5B & 14.03 & 26.73 & 33.29 & 25.41 & \textbf{80.44} & \textbf{93.88} & 7.45 & 18.59 & 38.78 & 23.24 & \textbf{84.52} & \textbf{94.90} \\
        \midrule
        Qwen-Audio-3.0-TTS & \textbf{0.30} & \textbf{0.31} & \underline{5.62} & \underline{2.22} & \underline{78.85} & \underline{93.16} & 3.30 & \underline{6.72} & \underline{4.85} & 5.00 & \underline{82.35} & \underline{93.45} \\
        \bottomrule
    \end{tabular}}
\end{table}

We evaluate one-pass synthesis without external segmentation or audio stitching. Content fidelity is measured by CER/WER, P-SIM measures similarity to the prompt, and S-SIM measures consistency among segments of the same generated utterance.


Table~\ref{tab:long-form-tts-full} shows that Qwen-Audio-3.0-TTS maintains competitive content accuracy and strong speaker consistency in both Chinese and English during one-pass long-form synthesis. It substantially improves content fidelity over CosyVoice3-1.5B while retaining high prompt and segment-level speaker similarity. The Chinese and English test sets each contain 100 paragraph-level inputs. Samples are divided into short, mid, and long buckets by input length. P-SIM is the average similarity between prompt and generated segments, while S-SIM is the average pairwise similarity among overlapping segments within a generated utterance. Table~\ref{tab:long-form-tts-full} reports the complete bucket-level breakdown.

\subsubsection{Results of Acoustic Robustness}

This benchmark evaluates voice-clone robustness using real-world noisy, reverberant, and unclear enrollment speech. The three subsets respectively cover background interference, far-field or room reverberation, and predominantly telephone-like narrow-band speech with audible distortion. Unlike benchmarks based on synthetic corruption, these recordings have no paired clean references. Speaker similarity should therefore be viewed as an auxiliary measure of how well speaker cues are retained from degraded prompts, rather than as an absolute estimate.

\begin{table*}[htb]
    \caption{Objective zero-shot TTS results under noisy, reverberant, and unclear prompt conditions. Models marked with ``Denoise'' use their inference-time denoising mode. Content error, ERes2Net speaker similarity (SIM), and DNSMOS are reported. Bold and underlined values denote the best and second-best results in each column.}
    \vspace{6pt}
    \label{tab:acoustic-robustness}
    \centering
    \setlength\tabcolsep{1.2pt}
    \resizebox{\textwidth}{!}{
    \begin{tabular}{lccccccccc}
        \toprule
        \multirow{2}{*}{\textbf{Model}} 
        & \multicolumn{3}{c}{\textbf{Noisy}} 
        & \multicolumn{3}{c}{\textbf{Reverb}} 
        & \multicolumn{3}{c}{\textbf{Unclear}} \\
        \cmidrule(r){2-4} \cmidrule(r){5-7} \cmidrule(r){8-10}
        & \textbf{WER (\%)~$\downarrow$} & \textbf{SIM (\%)~$\uparrow$} & \textbf{DNSMOS~$\uparrow$}
        & \textbf{WER (\%)~$\downarrow$} & \textbf{SIM (\%)~$\uparrow$} & \textbf{DNSMOS~$\uparrow$}
        & \textbf{WER (\%)~$\downarrow$} & \textbf{SIM (\%)~$\uparrow$} & \textbf{DNSMOS~$\uparrow$} \\
        \midrule
        \multicolumn{10}{l}{\textbf{Commercial API Models}} \\

        MiniMax-Speech-2.8-HD & \underline{0.85} & 66.72 & 3.464 & 0.83 & 61.56 & 3.065 & \textbf{1.28} & 68.33 & 3.174 \\
        MiniMax-Speech-2.8-HD$\cdot$Denoise & \textbf{0.83} & 63.83 & 3.728 & 0.87 & 56.53 & 3.343 & 1.58 & 67.84 & 3.241 \\
        ElevenLabs-v3 & 1.17 & 46.91 & 3.779 & 1.75 & 41.46 & 3.090 & 1.67 & 47.12 & 3.304 \\
        ElevenLabs-v3$\cdot$Denoise & 1.19 & 46.07 & \textbf{3.981} & \textbf{0.58} & 44.39 & \textbf{4.025} & \underline{1.38} & 43.90 & \textbf{3.496} \\

        \midrule
        \multicolumn{10}{l}{\textbf{Open-source Models}} \\
        
        Qwen3-TTS-12Hz-1.7B-Base & 2.01 & 65.61 & 3.595 & 2.11 & 63.42 & 2.887 & 2.85 & 70.62 & 3.050 \\
        LongCat-AudioDiT-3.5B & 3.44 & 58.70 & 3.777 & 1.05 & 56.91 & 3.169 & 3.12 & 73.54 & 3.262 \\
        Dots.TTS-2B (SOAR) & 2.90 & \textbf{76.69} & 3.221 & 2.12 & \underline{72.38} & 2.888 & 2.16 & \textbf{76.69} & 3.070 \\
        VoxCPM2$\cdot$Denoise & 4.36 & 65.31 & 3.678 & 10.07 & 51.05 & 2.830 & 6.71 & 68.81 & 3.051 \\
        CosyVoice3-1.5B & 1.56 & 75.40 & 3.301 & 1.78 & 71.91 & 3.021 & 2.39 & 72.06 & 3.113 \\
        \midrule
        \textbf{Qwen-Audio-3.0-TTS} & 1.18 & \underline{76.14} & \underline{3.962} & \underline{0.69} & \textbf{74.12} & \underline{3.925} & 1.61 & \underline{76.53} & \underline{3.305} \\
        \bottomrule
    \end{tabular}}
\end{table*}

Qwen-Audio-3.0-TTS is designed with built-in robustness to degraded enrollment speech, without relying on a dedicated inference-time denoising mode. As shown in Table~\ref{tab:acoustic-robustness}, this capability leads to strong results across all three conditions. We evaluate both standard and denoising modes for MiniMax-Speech-2.8-HD and ElevenLabs-v3. Enabling denoising raises MiniMax's DNSMOS from 3.464 to 3.728 on Noisy and from 3.065 to 3.343 on Reverb, but reduces SIM from 66.72 to 63.83 and from 61.56 to 56.53, respectively. A similar trade-off appears for ElevenLabs-v3, whose denoising mode improves Reverb DNSMOS from 3.090 to 4.025 and WER from 1.75\% to 0.58\%, while its SIM remains low at 44.39\%. By comparison, Qwen-Audio-3.0-TTS reaches DNSMOS scores of 3.962 and 3.925 on Noisy and Reverb, close to ElevenLabs-v3$\cdot$Denoise, while achieving much higher SIM scores of 76.14\% and 74.12\%. On Reverb, it further obtains the best SIM together with the second-best WER and DNSMOS, showing a strong balance among denoising quality, intelligibility, and speaker preservation.

\subsubsection{Results of Instruction Following under Zero-shot Voice Cloning}

The benchmark contains 440 zero-shot voice-cloning cases, evenly split between Chinese and English, with single-attribute and natural-language or structured multi-attribute instructions. For speaker preservation, ERes2Net cosine similarity is computed between the prompt and synthesized utterance. Gemini-2.5-Pro~\cite{comanici2025gemini25} is instructed to judge only how the utterance is spoken. Single-attribute cases receive a binary score. Complex instructions are decomposed into affect, rate, volume, clarity, rhythm, and intonation; an LLM selects the three most important dimensions for each instruction and instantiates audible criteria before any system is evaluated. Each criterion is binary, giving a score from 0 to 3, and the criteria remain fixed across systems.

On a stratified calibration subset, Gemini reaches 70.0\% agreement with human judgments for single-attribute instructions, with 92.3\% precision, 64.9\% recall, and 76.2\% F1. Its lower positive rate indicates conservative judging (McNemar's test, $p=0.007$). For complex instructions, criterion-level agreement is 56.7\%, 72.0\% of final scores differ by at most one point, and mean scores are 1.46 for humans and 1.48 for Gemini. Blind review finds 35.4\% of disputed criteria inherently ambiguous, with both judgments defensible.

\begin{table*}[t]
    \caption{Instruction-following performance and speaker similarity (\%) on
    the bilingual benchmark. \textnormal{SA}, \textnormal{NL}, and
    \textnormal{ST} denote single-attribute, natural-language multi-attribute,
    and structured-attribute instructions, respectively. Speaker similarity
    is measured by the ERes2Net cosine similarity between each synthesized
    utterance and its reference utterance. Overall is computed by averaging over all evaluation instances pooled across subsets. Bold and underlined values indicate
    the best and second-best results in each column, respectively.}
    \label{tab:bilingual_instruction_following}
    \centering
    \resizebox{\textwidth}{!}{%
        \begin{tabular}{lcccccccccc}
        \toprule
        \multirow[c]{2}{*}{\textbf{Model}}
        & \multicolumn{4}{c}{\textbf{Instruction Following (zh)}}
        & \multicolumn{4}{c}{\textbf{Instruction Following (en)}}
        & \multicolumn{2}{c}{\textbf{Speaker Similarity}} \\
        \cmidrule(lr){2-5}
        \cmidrule(lr){6-9}
        \cmidrule(lr){10-11}
    
        & \textbf{SA} & \textbf{NL} & \textbf{ST} & \textbf{Overall}
        & \textbf{SA} & \textbf{NL} & \textbf{ST} & \textbf{Overall}
        & \textbf{zh} & \textbf{en} \\
        \midrule

        IndexTTS2
        & 65.00 & 42.67 & 40.67 & 54.39
        & \underline{67.50} & 37.33 & 62.00 & 59.39
        & 64.97 & 66.25 \\
        
        CosyVoice3-1.5B
        & \underline{82.50} & \underline{67.33} & \textbf{68.67} & \underline{75.91}
        & 63.33 & \underline{56.00} & \underline{74.00} & \underline{64.09}
        & \textbf{75.60} & \textbf{68.45} \\
        
        \midrule
        
        Qwen-Audio-3.0-TTS
        & \textbf{87.50} & \textbf{72.00} & \underline{65.33}
        & \textbf{78.94}
        & \textbf{83.33} & \textbf{76.00} & \textbf{78.00}
        & \textbf{80.45}
        & \underline{73.27} & \underline{66.66} \\
        \bottomrule
        \end{tabular}%
    }
\end{table*}

We compare Qwen-Audio-3.0-TTS with IndexTTS2~\cite{zhou2025indextts2breakthroughemotionallyexpressive} and CosyVoice3-1.5B, two recent systems that support style-controlled synthesis conditioned on a reference utterance. As shown in Table~\ref{tab:bilingual_instruction_following}, Qwen-Audio-3.0-TTS obtains the best overall instruction-following results in both \textnormal{zh} and \textnormal{en}, with scores of 78.94 and 80.45, respectively. Its advantage is especially clear for natural-language instructions, suggesting a stronger ability to interpret flexible descriptions and translate multiple style requirements into audible speech characteristics. Compared with CosyVoice3-1.5B, it improves bilingual instruction following while maintaining competitive speaker similarity, demonstrating stronger style control without substantially compromising prompt-voice preservation.

\subsection{Objective Results on Speaker-Adapted Models}

\begin{figure}[t]
    \centering
    \includegraphics[width=0.92\textwidth]{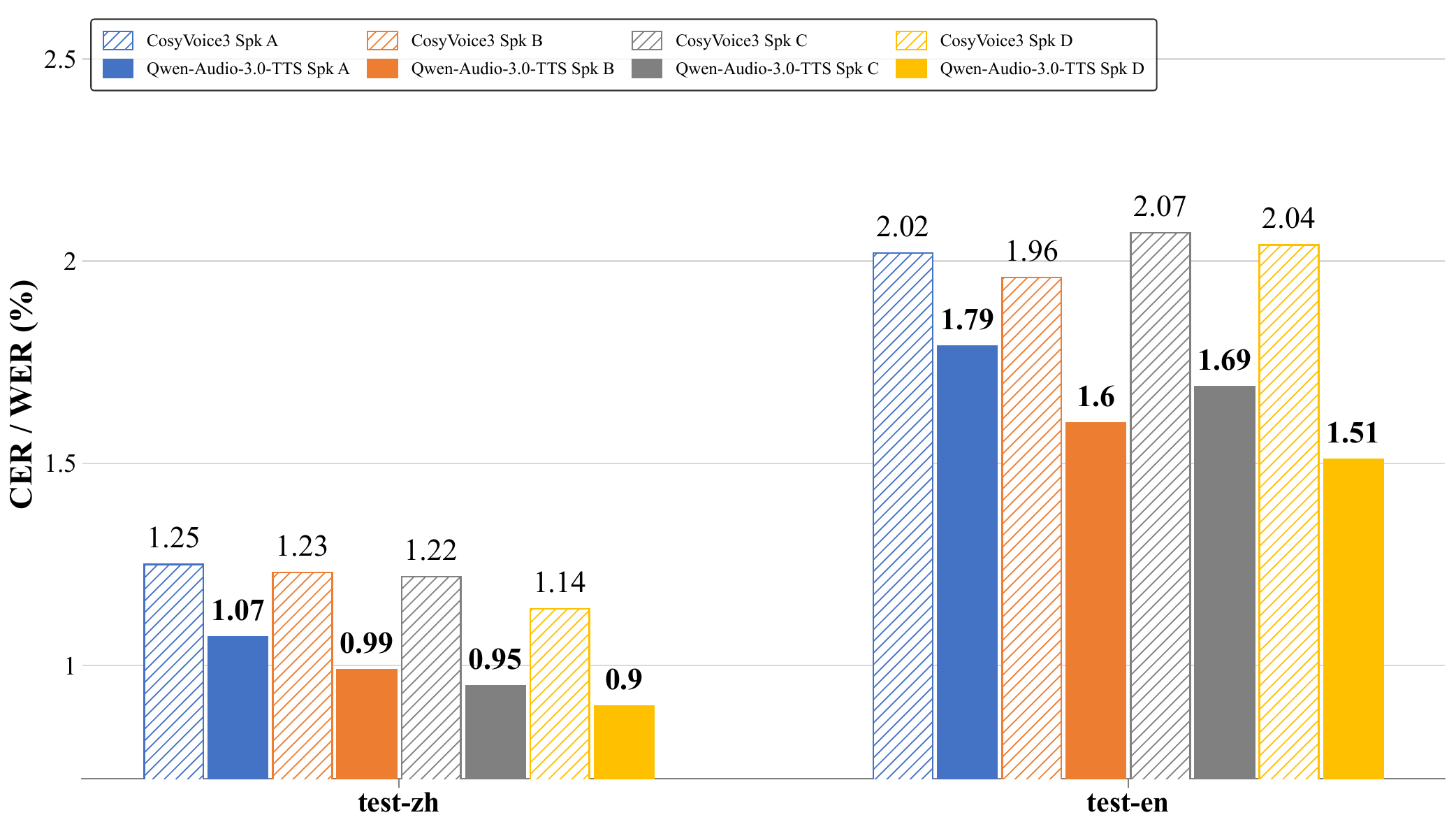}
    \caption{Content-consistency results for speaker-adapted CosyVoice3 and Qwen-Audio-3.0-TTS models on four anonymized speakers. Lower CER/WER is better.}
    \label{fig:sft-speakers}
\end{figure}

We compare speaker-adapted versions of CosyVoice3 and Qwen-Audio-3.0-TTS on the standard SEED-TTS-Eval \emph{test-zh} and \emph{test-en} sets for four anonymized target speakers. As shown in Figure~\ref{fig:sft-speakers}, Qwen-Audio-3.0-TTS consistently improves content consistency across all four speakers. On \emph{test-zh}, CER decreases from 1.25\% to 1.07\% for Speaker~A, from 1.23\% to 0.99\% for Speaker~B, from 1.22\% to 0.95\% for Speaker~C, and from 1.14\% to 0.90\% for Speaker~D. On \emph{test-en}, WER decreases from 2.02\% to 1.79\%, from 1.96\% to 1.60\%, from 2.07\% to 1.69\%, and from 2.04\% to 1.51\%, respectively.

These CER/WER metrics measure content consistency and help detect intelligibility and pronunciation regressions.

\subsection{Subjective Evaluation Results}
\subsubsection{Subjective Evaluation on Dialect Speech Synthesis}

We conduct a native-speaker subjective evaluation covering 20 Chinese dialects with 50 prompts per dialect. Three native speakers independently annotate every utterance. \textbf{Dialect Authenticity} measures whether speech achieves the requested regional variety without drifting toward Mandarin or another dialect. \textbf{Pronunciation accuracy} penalizes missing, substituted, and inserted characters. \textbf{Prosodic naturalness} assesses pace, pause placement, and intonation against native habits. Annotators use four severity levels---\emph{Perfect}, \emph{P2}, \emph{P1}, and \emph{P0}---mapped to scores from 4 to 1, plus diagnostic error tags.

For aggregate analysis, we map the four severity levels to numeric scores: \emph{Perfect}=4, \emph{P2}=3, \emph{P1}=2, and \emph{P0}=1. The mean score is then used as a dimension-level perceptual quality indicator, where a higher score indicates better subjective quality.

\begin{table}[htb]
    \caption{Subjective evaluation results of Qwen-Audio-3.0-TTS on
    multi-dialect speech synthesis. Each dimension is rated with four severity
    levels, where \emph{Perfect}, \emph{P2}, \emph{P1}, and \emph{P0} are
    mapped to scores of 4, 3, 2, and 1, respectively.}
    \label{tab:dialect-subjective}
    \centering
    \small
    \begin{tabular}{lccccc}
        \toprule
        \textbf{Dimension} & \textbf{Perfect} & \textbf{P2} & \textbf{P1}
        & \textbf{P0} & \textbf{Mean Score} \\
        \midrule
        \textbf{Dialect Authenticity}   & 66.7\% & 31.5\% & 0.8\% & 1.0\% & 3.639 \\
        \textbf{Pronunciation Accuracy} & 93.5\% & 6.5\%  & 0.0\% & 0.0\% & 3.935 \\
        \textbf{Prosodic Naturalness}   & 68.1\% & 31.9\% & 0.0\% & 0.0\% & 3.680 \\
        \bottomrule
    \end{tabular}
\end{table}

As shown in Table~\ref{tab:dialect-subjective}, Qwen-Audio-3.0-TTS achieves strong overall intelligibility and perceptual quality in multi-dialect speech synthesis. Pronunciation accuracy is the most stable dimension, with 93.5\% \emph{Perfect} labels and no \emph{P1}/\emph{P0} cases, suggesting reliable character-level content preservation across dialects. Dialect authenticity and prosodic naturalness obtain 66.7\% and 68.1\% \emph{Perfect} while most remaining cases are mild \emph{P2} errors.

\subsubsection{Subjective Evaluation of Instruction Following Capabilities}

We conduct Arena-based human evaluation to assess the instruction-controlled speech synthesis capability of Qwen-Audio-3.0-TTS. The evaluation considers two independently rated dimensions: \emph{Instruction Following} and \emph{Prosodic Naturalness}. Instruction Following measures whether the synthesized speech accurately follows the specified control instructions, including emotion, speaking rate, role, and speaking style. Prosodic Naturalness evaluates whether the generated prosody sounds natural and human-like, regardless of whether the target instruction is correctly satisfied. This separation allows us to distinguish instruction controllability from perceptual naturalness.

The evaluation results are shown in Table~\ref{tab:arena_human_eval}. Qwen-Audio-3.0-TTS achieves win rates of 44.8\% and 55.6\% on the two dimensions, clearly improving over the previous-generation baseline.

\begin{table}[htb]
\caption{Arena-based human evaluation of instruction-controlled synthesis. Win rates are reported in percentage; higher is better, and the best result in each column is shown in bold.}
\label{tab:arena_human_eval}
\centering
\small
\begin{tabular}{lcc}
\toprule
\textbf{System} & \textbf{Instruction Following} & \textbf{Prosodic Naturalness} \\
\midrule
Previous-Gen Baseline  & 30.9 & 42.9 \\
Qwen-Audio-3.0-TTS  & \textbf{44.8} & \textbf{55.6} \\
\bottomrule
\end{tabular}
\end{table}

\section{Conclusion}

In this report, we present Qwen-Audio-3.0-TTS, a multilingual, freely controllable and highly robust speech synthesis system oriented towards production deployment. Its 12.5~Hz speech tokenizer, together with chunk-based flow-matching and causal vocoder, reduces end-to-end latency. Joint LM--FM training and LM reinforcement learning deliver a strong balance of content consistency and naturalness. Acoustic robustness training and FM reinforcement learning progressively improve speaker similarity, audio fidelity and robustness. Qwen-Audio-3.0-TTS achieves state-of-the-art performance on many reported dimensions or the strongest aggregate scores across SEED-TTS-Eval, CV3-Eval, instruction-following, long-form, and adverse-prompt evaluations. It also ranks first on the independent Artificial Analysis Text-to-Speech Arena leaderboard released on July 16, 2026.

Beyond individual benchmark gains, Qwen-Audio-3.0-TTS advances TTS toward a unified and practical speech-generation system. A single model integrates multilingual and multi-dialect synthesis, zero-shot voice cloning, free-style instruction following, fine-grained inline control, one-pass long-form generation and robustness to degraded real-world prompts. In addition, we propose a scalable speaker adaptation protocol based on Qwen-Audio-3.0-TTS. We believe this provides a practical foundation for the next generation of general-purpose, controllable, and deployment-ready TTS systems. 

\section*{Acknowledgements}

We thank Bangduo Chen, Biao Tian, Bin Ma, Bin Yuan, Binbin Zhang, Chaohong Tan, Chen Ding, Chong Deng, Chongde Zhang, Gang Qiao, Hongzhi Cai, Jianwei Yu, Jiaqi Shi, Jiaqing Liu, Jixing Yu, Junhao Xu, Lingyun Zuo, Menglin Wu, Qian Chen, Sitong Zhao, Xian Yang, Yajie Wen, Yang Bai, Yiping Peng, Yuting Teng, Ze Xu, Zhenglin Wang, Zhifu Gao, Ziyi Cheng for their valuable
contributions to data curation, system development, evaluation, and infrastructure support. Names
are listed alphabetically by given name.


\clearpage
\bibliographystyle{unsrt}
\bibliography{ref}

@article{xu2025qwen2,
  title={Qwen2.5-omni technical report},
  author={Xu, Jin and Guo, Zhifang and He, Jinzheng and Hu, Hangrui and He, Ting and Bai, Shuai and Chen, Keqin and Wang, Jialin and Fan, Yang and Dang, Kai and others},
  journal={arXiv preprint arXiv:2503.20215},
  year={2025}
}

@article{park2024librispeechlong,
  author       = {Se Jin Park and
                  Julian Salazar and
                  Aren Jansen and
                  Keisuke Kinoshita and
                  Yong Man Ro and
                  R. J. Skerry{-}Ryan},
  title        = {Long-Form Speech Generation with Spoken Language Models},
  journal      = {CoRR},
  volume       = {abs/2412.18603},
  year         = {2024}
}

@misc{xie2025fireredtts2,
      title={FireRedTTS-2: Towards Long Conversational Speech Generation for Podcast and Chatbot}, 
      author={Kun Xie and Feiyu Shen and Junjie Li and Fenglong Xie and Xu Tang and Yao Hu},
      year={2025},
      eprint={2509.02020},
      archivePrefix={arXiv},
      primaryClass={cs.SD},
      url={https://arxiv.org/abs/2509.02020}, 
}

@misc{qwenteam2026qwen35omni,
      title={Qwen3.5-Omni Technical Report}, 
      author={Qwen Team},
      year={2026},
      eprint={2604.15804},
      archivePrefix={arXiv},
      primaryClass={cs.CL},
      url={https://arxiv.org/abs/2604.15804}, 
}

@article{jia2025ditar,
  title={DiTAR: Diffusion Transformer Autoregressive Modeling for Speech Generation},
  author={Jia, Dongya and Chen, Zhuo and Chen, Jiawei and Du, Chenpeng and Wu, Jian and Cong, Jian and Zhuang, Xiaobin and Li, Chumin and Wei, Zhen and Wang, Yuping and others},
  journal={arXiv preprint arXiv:2502.03930},
  year={2025}
}

@article{wang2025spark,
  title={Spark-tts: An efficient llm-based text-to-speech model with single-stream decoupled speech tokens},
  author={Wang, Xinsheng and Jiang, Mingqi and Ma, Ziyang and Zhang, Ziyu and Liu, Songxiang and Li, Linqin and Liang, Zheng and Zheng, Qixi and Wang, Rui and Feng, Xiaoqin and others},
  journal={arXiv preprint arXiv:2503.01710},
  year={2025}
}

@article{du2024cosyvoice2,
  title={CosyVoice 2: Scalable Streaming Speech Synthesis with Large Language Models},
  author={Du, Zhihao and Wang, Yuxuan and Chen, Qian and Shi, Xian and Lv, Xiang and Zhao, Tianyu and Gao, Zhifu and Yang, Yexin and Gao, Changfeng and Wang, Hui and others},
  journal={arXiv preprint arXiv:2412.10117},
  year={2024}
}

@inproceedings{DBLP:conf/icassp/ReddyGC22,
	author       = {Chandan K. A. Reddy and
	Vishak Gopal and
	Ross Cutler},
	title        = {Dnsmos {P.835:} {A} Non-Intrusive Perceptual Objective Speech Quality
	Metric to Evaluate Noise Suppressors},
	booktitle    = {{ICASSP}},
	pages        = {886--890},
	publisher    = {{IEEE}},
	year         = {2022}
}

@article{DBLP:journals/corr/abs-2406-02430,
	author       = {Philip Anastassiou and
	Jiawei Chen and
	Jitong Chen and
	Yuanzhe Chen and
	Zhuo Chen and
	Ziyi Chen and
	Jian Cong and
	Lelai Deng and
	Chuang Ding and
	Lu Gao and
	Mingqing Gong and
	Peisong Huang and
	Qingqing Huang and
	Zhiying Huang and
	Yuanyuan Huo and
	Dongya Jia and
	Chumin Li and
	Feiya Li and
	Hui Li and
	Jiaxin Li and
	Xiaoyang Li and
	Xingxing Li and
	Lin Liu and
	Shouda Liu and
	Sichao Liu and
	Xudong Liu and
	Yuchen Liu and
	Zhengxi Liu and
	Lu Lu and
	Junjie Pan and
	Xin Wang and
	Yuping Wang and
	Yuxuan Wang and
	Zhen Wei and
	Jian Wu and
	Chao Yao and
	Yifeng Yang and
	Yuanhao Yi and
	Junteng Zhang and
	Qidi Zhang and
	Shuo Zhang and
	Wenjie Zhang and
	Yang Zhang and
	Zilin Zhao and
	Dejian Zhong and
	Xiaobin Zhuang},
	title        = {Seed-TTS: {A} Family of High-Quality Versatile Speech Generation Models},
	journal      = {CoRR},
	volume       = {abs/2406.02430},
	year         = {2024}
}

@article{DBLP:journals/corr/abs-2410-06885,
	author       = {Yushen Chen and
	Zhikang Niu and
	Ziyang Ma and
	Keqi Deng and
	Chunhui Wang and
	Jian Zhao and
	Kai Yu and
	Xie Chen},
	title        = {{F5-TTS:} {A} Fairytaler that Fakes Fluent and Faithful Speech with
	Flow Matching},
	journal      = {CoRR},
	volume       = {abs/2410.06885},
	year         = {2024}
}

@article{DBLP:journals/corr/abs-2406-18009,
	author       = {Sefik Emre Eskimez and
	Xiaofei Wang and
	Manthan Thakker and
	Canrun Li and
	Chung{-}Hsien Tsai and
	Zhen Xiao and
	Hemin Yang and
	Zirun Zhu and
	Min Tang and
	Xu Tan and
	Yanqing Liu and
	Sheng Zhao and
	Naoyuki Kanda},
	title        = {{E2} {TTS:} Embarrassingly Easy Fully Non-Autoregressive Zero-Shot
	{TTS}},
	journal      = {CoRR},
	volume       = {abs/2406.18009},
	year         = {2024}
}

@inproceedings{DBLP:conf/nips/LeVSKSMWMAMH23,
	author       = {Matthew Le and
	Apoorv Vyas and
	Bowen Shi and
	Brian Karrer and
	Leda Sari and
	Rashel Moritz and
	Mary Williamson and
	Vimal Manohar and
	Yossi Adi and
	Jay Mahadeokar and
	Wei{-}Ning Hsu},
	title        = {Voicebox: Text-Guided Multilingual Universal Speech Generation at
	Scale},
	booktitle    = {NeurIPS},
	year         = {2023}
}

@article{sun2025f5r,
  title={F5R-TTS: Improving Flow Matching based Text-to-Speech with Group Relative Policy Optimization},
  author={Sun, Xiaohui and Xiao, Ruitong and Mo, Jianye and Wu, Bowen and Yu, Qun and Wang, Baoxun},
  journal={arXiv preprint arXiv:2504.02407},
  year={2025}
}

@inproceedings{chen2022large,
  title={Large-scale self-supervised speech representation learning for automatic speaker verification},
  author={Chen, Zhengyang and Chen, Sanyuan and Wu, Yu and Qian, Yao and Wang, Chengyi and Liu, Shujie and Qian, Yanmin and Zeng, Michael},
  booktitle={ICASSP 2022-2022 IEEE International Conference on Acoustics, Speech and Signal Processing (ICASSP)},
  pages={6147--6151},
  year={2022},
  organization={IEEE}
}

@article{cosyvoice,
	author       = {Zhihao Du and
	Qian Chen and
	Shiliang Zhang and
	Kai Hu and
	Heng Lu and
	Yexin Yang and
	Hangrui Hu and
	Siqi Zheng and
	Yue Gu and
	Ziyang Ma and
	Zhifu Gao and
	Zhijie Yan},
	title        = {CosyVoice: {A} Scalable Multilingual Zero-shot Text-to-speech Synthesizer
	based on Supervised Semantic Tokens},
	journal      = {CoRR},
	volume       = {abs/2407.05407},
	year         = {2024}
}

@misc{qwen2.5,
	title = {Qwen2.5: A Party of Foundation Models},
	url = {https://qwenlm.github.io/blog/qwen2.5/},
	author = {{Qwen Team}},
	month = {September},
	year = {2024}
}

@inproceedings{DBLP:conf/iclr/MentzerMAT24,
	author       = {Fabian Mentzer and
	David Minnen and
	Eirikur Agustsson and
	Michael Tschannen},
	title        = {Finite Scalar Quantization: {VQ-VAE} Made Simple},
	booktitle    = {{ICLR}},
	publisher    = {OpenReview.net},
	year         = {2024}
}

@article{DBLP:journals/ijon/SuALPBL24,
	author       = {Jianlin Su and
	Murtadha H. M. Ahmed and
	Yu Lu and
	Shengfeng Pan and
	Wen Bo and
	Yunfeng Liu},
	title        = {RoFormer: Enhanced transformer with Rotary Position Embedding},
	journal      = {Neurocomputing},
	volume       = {568},
	pages        = {127063},
	year         = {2024}
}

@article{DBLP:journals/corr/abs-2207-12598,
  author       = {Jonathan Ho and
                  Tim Salimans},
  title        = {Classifier-Free Diffusion Guidance},
  journal      = {CoRR},
  volume       = {abs/2207.12598},
  year         = {2022},
  url          = {https://doi.org/10.48550/arXiv.2207.12598},
  doi          = {10.48550/ARXIV.2207.12598},
  eprinttype    = {arXiv},
  eprint       = {2207.12598},
  bibsource    = {dblp computer science bibliography, https://dblp.org}
}

@inproceedings{DBLP:conf/icml/RadfordKXBMS23,
  author       = {Alec Radford and
                  Jong Wook Kim and
                  Tao Xu and
                  Greg Brockman and
                  Christine McLeavey and
                  Ilya Sutskever},
  editor       = {Andreas Krause and
                  Emma Brunskill and
                  Kyunghyun Cho and
                  Barbara Engelhardt and
                  Sivan Sabato and
                  Jonathan Scarlett},
  title        = {Robust Speech Recognition via Large-Scale Weak Supervision},
  booktitle    = {International Conference on Machine Learning, {ICML} 2023, 23-29 July
                  2023, Honolulu, Hawaii, {USA}},
  series       = {Proceedings of Machine Learning Research},
  volume       = {202},
  pages        = {28492--28518},
  publisher    = {{PMLR}},
  year         = {2023},
  url          = {https://proceedings.mlr.press/v202/radford23a.html},
  bibsource    = {dblp computer science bibliography, https://dblp.org}
}

@article{DBLP:journals/corr/abs-2301-02111,
  author       = {Chengyi Wang and
                  Sanyuan Chen and
                  Yu Wu and
                  Ziqiang Zhang and
                  Long Zhou and
                  Shujie Liu and
                  Zhuo Chen and
                  Yanqing Liu and
                  Huaming Wang and
                  Jinyu Li and
                  Lei He and
                  Sheng Zhao and
                  Furu Wei},
  title        = {Neural Codec Language Models are Zero-Shot Text to Speech Synthesizers},
  journal      = {CoRR},
  volume       = {abs/2301.02111},
  year         = {2023},
  url          = {https://doi.org/10.48550/arXiv.2301.02111},
  doi          = {10.48550/ARXIV.2301.02111},
  eprinttype    = {arXiv},
  eprint       = {2301.02111},
  bibsource    = {dblp computer science bibliography, https://dblp.org}
}

@inproceedings{DBLP:conf/interspeech/GaoZ0Y22,
	author       = {Zhifu Gao and
	Shiliang Zhang and
	Ian McLoughlin and
	Zhijie Yan},
	title        = {Paraformer: Fast and Accurate Parallel Transformer for Non-autoregressive
	End-to-End Speech Recognition},
	booktitle    = {{Interspeech}},
	pages        = {2063--2067},
	publisher    = {{ISCA}},
	year         = {2022}
}

@article{chen2023enhanced,
  title={An enhanced res2net with local and global feature fusion for speaker verification},
  author={Chen, Yafeng and Zheng, Siqi and Wang, Hui and Cheng, Luyao and Chen, Qian and Qi, Jiajun},
  journal={arXiv preprint arXiv:2305.12838},
  year={2023}
}

@article{gao2023funasr,
  title={Funasr: A fundamental end-to-end speech recognition toolkit},
  author={Gao, Zhifu and Li, Zerui and Wang, Jiaming and Luo, Haoneng and Shi, Xian and Chen, Mengzhe and Li, Yabin and Zuo, Lingyun and Du, Zhihao and Xiao, Zhangyu and others},
  journal={arXiv preprint arXiv:2305.11013},
  year={2023}
}

@article{chen2025minmo,
  title={{MinMo}: A Multimodal Large Language Model for Seamless Voice Interaction},
  author={Chen, Qian and Chen, Yafeng and Chen, Yanni and Chen, Mengzhe and Chen, Yingda and Deng, Chong and Du, Zhihao and Gao, Ruize and Gao, Changfeng and Gao, Zhifu and others},
  journal={arXiv preprint arXiv:2501.06282},
  year={2025}
}

@misc{funaduiollm,
  title={FunAudioLLM: Voice Understanding and Generation Foundation Models for Natural Interaction Between Humans and LLMs},
  author={Keyu An and Qian Chen and Chong Deng and Zhihao Du and Changfeng Gao and Zhifu Gao and Yue Gu and Ting He and Hangrui Hu and Kai Hu and Shengpeng Ji and Yabin Li and Zerui Li and Heng Lu and Haoneng Luo and Xiang Lv and Bin Ma and Ziyang Ma and Chongjia Ni and Changhe Song and Jiaqi Shi and Xian Shi and Hao Wang and Wen Wang and Yuxuan Wang and Zhangyu Xiao and Zhijie Yan and Yexin Yang and Bin Zhang and Qinglin Zhang and Shiliang Zhang and Nan Zhao and Siqi Zheng},
  year={2024},
  eprint={2407.04051},
  archivePrefix={arXiv},
  primaryClass={cs.SD},
  url={https://arxiv.org/abs/2407.04051}
}

@inproceedings{fleurs,
  title={Fleurs: Few-shot learning evaluation of universal representations of speech},
  author={Conneau, Alexis and Ma, Min and Khanuja, Simran and Zhang, Yu and Axelrod, Vera and Dalmia, Siddharth and Riesa, Jason and Rivera, Clara and Bapna, Ankur},
  booktitle={2022 IEEE Spoken Language Technology Workshop (SLT)},
  pages={798--805},
  year={2023},
  organization={IEEE}
}

@article{commonvoice,
  title={Common voice: A massively-multilingual speech corpus},
  author={Ardila, Rosana and Branson, Megan and Davis, Kelly and Henretty, Michael and Kohler, Michael and Meyer, Josh and Morais, Reuben and Saunders, Lindsay and Tyers, Francis M and Weber, Gregor},
  journal={arXiv preprint arXiv:1912.06670},
  year={2019}
}

@misc{lian2026dotsttstechnicalreport,
      title={dots.tts Technical Report}, 
      author={Shi Lian and Changtao Li and Bohan Li and Hankun Wang and Da Zheng and Junfeng Tian and Yufeng Ma and Colin Zhang and Kai Yu},
      year={2026},
      eprint={2606.07080},
      archivePrefix={arXiv},
      primaryClass={cs.SD},
      url={https://arxiv.org/abs/2606.07080}, 
}

@misc{xin2026longcataudiodithighfidelitydiffusiontexttospeech,
      title={LongCat-AudioDiT: High-Fidelity Diffusion Text-to-Speech in the Waveform Latent Space}, 
      author={Detai Xin and Shujie Hu and Chengzuo Yang and Chen Huang and Guoqiao Yu and Guanglu Wan and Xunliang Cai},
      year={2026},
      eprint={2603.29339},
      archivePrefix={arXiv},
      primaryClass={cs.SD},
      url={https://arxiv.org/abs/2603.29339}, 
}

@misc{zhou2026voxcpm2technicalreport,
      title={VoxCPM2 Technical Report}, 
      author={Yixuan Zhou and Guoyang Zeng and Xin Liu and Xiang Li and Renjie Yu and Jiancheng Gui and Jiaheng Wu and Ziyang Wang and Xudong Shen and Runchuan Ye and Zhisheng Zhang and Jiuyang Zhou and Bingsong Bai and Weiyue Sun and Mengyuan Deng and Qundong Shi and Zhiyong Wu and Zhiyuan Liu},
      year={2026},
      eprint={2606.06928},
      archivePrefix={arXiv},
      primaryClass={cs.SD},
      url={https://arxiv.org/abs/2606.06928}, 
}

@misc{du2025cosyvoice3inthewildspeech,
      title={CosyVoice 3: Towards In-the-wild Speech Generation via Scaling-up and Post-training}, 
      author={Zhihao Du and Changfeng Gao and Yuxuan Wang and Fan Yu and Tianyu Zhao and Hao Wang and Xiang Lv and Hui Wang and Chongjia Ni and Xian Shi and Keyu An and Guanrou Yang and Yabin Li and Yanni Chen and Zhifu Gao and Qian Chen and Yue Gu and Mengzhe Chen and Yafeng Chen and Shiliang Zhang and Wen Wang and Jieping Ye},
      year={2025},
      eprint={2505.17589},
      archivePrefix={arXiv},
      primaryClass={cs.SD},
      url={https://arxiv.org/abs/2505.17589}, 
}

@misc{yu2025joyvoice,
      title={JoyVoice: Long-Context Conditioning for Anthropomorphic Multi-Speaker Conversational Synthesis},
      author={Fan Yu and Tao Wang and You Wu and Lin Zhu and Wei Deng and Weisheng Han and Wenchao Wang and Lin Hu and Xiangyu Liang and Xiaodong He and Yankun Huang and Yu Gu and Yuan Liu and Yuxuan Wang and Zhangyu Xiao and Ziteng Wang and Boya Dong and Feng Dang and Jinming Chen and Jingdong Li and Jun Wang and Yechen Jin and Yuan Zhang and Zhengyan Sheng and Xin Wang},
      year={2025},
      eprint={2512.19090},
      archivePrefix={arXiv},
      primaryClass={cs.SD},
      url={https://arxiv.org/abs/2512.19090},
}

@misc{hu2026qwen3ttstechnicalreport,
      title={Qwen3-TTS Technical Report}, 
      author={Hangrui Hu and Xinfa Zhu and Ting He and Dake Guo and Bin Zhang and Xiong Wang and Zhifang Guo and Ziyue Jiang and Hongkun Hao and Zishan Guo and Xinyu Zhang and Pei Zhang and Baosong Yang and Jin Xu and Jingren Zhou and Junyang Lin},
      year={2026},
      eprint={2601.15621},
      archivePrefix={arXiv},
      primaryClass={cs.SD},
      url={https://arxiv.org/abs/2601.15621}, 
}

@misc{minimax2025minimaxspeechintrinsiczeroshottexttospeech,
  title         = {MiniMax-Speech: Intrinsic Zero-Shot Text-to-Speech with a Learnable Speaker Encoder},
  author        = {Bowen Zhang and Congchao Guo and Geng Yang and Hang Yu and
                   Haozhe Zhang and Heidi Lei and Jialong Mai and Junjie Yan and
                   Kaiyue Yang and Mingqi Yang and Peikai Huang and Ruiyang Jin and
                   Sitan Jiang and Weihua Cheng and Yawei Li and Yichen Xiao and
                   Yiying Zhou and Yongmao Zhang and Yuan Lu and Yucen He},
  year          = {2025},
  eprint        = {2505.07916},
  archivePrefix = {arXiv},
  primaryClass  = {eess.AS},
  url           = {https://arxiv.org/abs/2505.07916}
}

@misc{zhou2025indextts2breakthroughemotionallyexpressive,
      title={IndexTTS2: A Breakthrough in Emotionally Expressive and Duration-Controlled Auto-Regressive Zero-Shot Text-to-Speech}, 
      author={Siyi Zhou and Yiquan Zhou and Yi He and Xun Zhou and Jinchao Wang and Wei Deng and Jingchen Shu},
      year={2025},
      eprint={2506.21619},
      archivePrefix={arXiv},
      primaryClass={cs.CL},
      url={https://arxiv.org/abs/2506.21619}, 
}

@article{flowgrpo,
  title={Flow-grpo: Training flow matching models via online rl},
  author={Liu, Jie and Liu, Gongye and Liang, Jiajun and Li, Yangguang and Liu, Jiaheng and Wang, Xintao and Wan, Pengfei and Zhang, Di and Ouyang, Wanli},
  journal={arXiv preprint arXiv:2505.05470},
  year={2025}
}

@INPROCEEDINGS{flowse_grpo,
  author={Wang, Haoxu and Tian, Biao and Jiang, Yiheng and Pan, Zexu and Zhao, Shengkui and Ma, Bin and Chen, Daren and Li, Xiangang},
  booktitle={ICASSP 2026 - 2026 IEEE International Conference on Acoustics, Speech and Signal Processing}, 
  title={FlowSE-GRPO: Training Flow Matching Speech Enhancement via Online Reinforcement Learning}, 
  year={2026},
  volume={},
  number={},
  pages={16182-16186},
  doi={10.1109/ICASSP55912.2026.11461623}}

@misc{flowtts_grpo,
      title={FlowTTS-GRPO: Online Reinforcement Learning with Multi-Objective Reward Optimization for Flow-Matching Based Text-to-Speech}, 
      author={Haoxu Wang and Biao Tian and Weiqin Li and Xiang Lv and Han Zhao and Xiangang Li},
      year={2026},
      eprint={2606.23190},
      archivePrefix={arXiv},
      primaryClass={eess.AS},
      url={https://arxiv.org/abs/2606.23190}, 
}

@inproceedings{lee2023bigvgan,
      title={BigVGAN: A Universal Neural Vocoder with Large-Scale Training},
      author={Sang-gil Lee and Wei Ping and Boris Ginsburg and Bryan Catanzaro and Sungroh Yoon},
      booktitle={The Eleventh International Conference on Learning Representations},
      year={2023},
      url={https://openreview.net/forum?id=iTtGCMDEzS_},
}

@misc{zhan2025vstyle,
      title={VStyle: A Benchmark for Voice Style Adaptation with Spoken Instructions},
      author={Jun Zhan and Mingyang Han and Yuxuan Xie and Chen Wang and Dong Zhang and Kexin Huang and Haoxiang Shi and DongXiao Wang and Tengtao Song and Qinyuan Cheng and Shimin Li and Jun Song and Xipeng Qiu and Bo Zheng},
      year={2025},
      eprint={2509.09716},
      archivePrefix={arXiv},
      primaryClass={cs.SD},
      url={https://arxiv.org/abs/2509.09716},
}

@inproceedings{hu2022lora,
      title={{LoRA}: Low-Rank Adaptation of Large Language Models},
      author={Edward J. Hu and Yelong Shen and Phillip Wallis and Zeyuan Allen-Zhu and Yuanzhi Li and Shean Wang and Lu Wang and Weizhu Chen},
      booktitle={International Conference on Learning Representations},
      year={2022},
      url={https://openreview.net/forum?id=nZeVKeeFYf9},
}

@misc{shao2024deepseekmath,
      title={DeepSeekMath: Pushing the Limits of Mathematical Reasoning in Open Language Models},
      author={Zhihong Shao and Peiyi Wang and Qihao Zhu and Runxin Xu and Junxiao Song and Xiao Bi and Haowei Zhang and Mingchuan Zhang and Y. K. Li and Y. Wu and Daya Guo},
      year={2024},
      eprint={2402.03300},
      archivePrefix={arXiv},
      primaryClass={cs.CL},
      url={https://arxiv.org/abs/2402.03300},
}

@inproceedings{jang2017categorical,
      title={Categorical Reparameterization with Gumbel-Softmax},
      author={Eric Jang and Shixiang Gu and Ben Poole},
      booktitle={International Conference on Learning Representations},
      year={2017},
      url={https://openreview.net/forum?id=rkE3y85ee},
}

@misc{comanici2025gemini25,
      title={Gemini 2.5: Pushing the Frontier with Advanced Reasoning, Multimodality, Long Context, and Next Generation Agentic Capabilities},
      author={Gheorghe Comanici and others},
      year={2025},
      eprint={2507.06261},
      archivePrefix={arXiv},
      primaryClass={cs.AI},
      url={https://arxiv.org/abs/2507.06261},
}

\end{document}